%% file: mnras_template.tex
\documentclass[fleqn,usenatbib]{mnras}

\usepackage[T1]{fontenc}
\usepackage{ae,aecompl}

\usepackage{graphicx}	
\usepackage{amsmath}	
\usepackage{float}
\usepackage{tabularx}
\usepackage[table,xcdraw]{xcolor}
\usepackage[normalem]{ulem}
\useunder{\uline}{\ul}{}
\usepackage{float}
\usepackage{amssymb}	
\usepackage{newtxtext,newtxmath}

\newcommand\eagle{{\sc EAGLE}}       
\newcommand\gadget{{\sc GADGET}}       
\newcommand\velociraptor{{\sc VELOCIraptor}}       
\newcommand\treefrog{{\sc TreeFrog}} 
\newcommand\chumm{{\sc chumm}} 

\title[The nature of gas accretion to haloes]{Revealing the physical properties of gas accreting to haloes in the EAGLE simulations}

\author[R. J. Wright et al.]{Ruby J. Wright\thanks{E-mail: ruby.wright@icrar.org}$^{ 1,2}$, Claudia del P. Lagos$^{1,2}$, Chris Power$^{1,2}$, Camila A. Correa$^3$
\\
$^{1}$International Centre for Radio Astronomy Research (ICRAR), University of Western Australia, Crawley, WA 6009, Australia\\
$^{2}$ARC Centre of Excellence for All Sky Astrophysics in 3 Dimensions (ASTRO 3D)\\
$^{3}$Institute for Theoretical Physics Amsterdam, University of Amsterdam, Science Park 904, 1098 XH Amsterdam, The Netherlands\\}

\date{Accepted XXX. Received YYY; in original form ZZZ}

\pubyear{2020}

\begin{document}
\label{firstpage}
\pagerange{\pageref{firstpage}--\pageref{lastpage}}
\maketitle

\begin{abstract}

The inflow of cosmological gas onto haloes, while challenging to directly observe and quantify, plays a fundamental role in the baryon cycle of galaxies. Using the \eagle\ suite of hydrodynamical simulations, we present a thorough exploration of the physical properties of gas accreting onto haloes -- namely, its spatial characteristics, density, temperature, and metallicity. Classifying accretion as ``hot'' or `` cold'' based on a temperature cut of $10^{5.5}{\rm K}$, we find that the covering fraction ($f_{\rm cov}$) of cold-mode accreting gas is significantly lower than the hot-mode, with $z=0$ $f_{\rm cov}$ values of $\approx 50\%$ and $\approx 80\%$ respectively. Active Galactic Nuclei (AGN) feedback in \eagle\ reduces inflow $f_{\rm cov}$ values by $\approx 10\%$, with outflows decreasing the solid angle available for accretion flows. Classifying inflow by particle history, we find that gas on first-infall onto a halo is metal-depleted by $\approx 2$~dex compared to pre-processed gas, which we find to mimic the circum-galactic medium (CGM) in terms of metal content. We also show that high (low) halo-scale gas accretion rates are associated with metal-poor (rich) CGM in haloes below $10^{12}M_{\odot}$, and that variation in halo-scale gas accretion rates may offer a physical explanation for the enhanced scatter in the star-forming main sequence at low ($\lesssim10^{9}M_{\odot}$) and high ($\gtrsim10^{10}M_{\odot}$) stellar masses. Our results highlight how gas inflow influences several halo- and galaxy-scale properties, and the need to combine kinematic and chemical data in order to confidently break the degeneracy between accreting and outgoing gas in CGM observations.

\end{abstract}




\section{Introduction}\label{sec:introduction}

The baryon cycle -- the interplay of gas inflow, outflow, and star formation at a galaxy and halo-level -- plays a dominant role in setting the  observable properties of galaxies over cosmic time. The accretion of gas onto dark matter haloes is a critical, yet a poorly understood aspect of the galactic baryon cycle. The necessity of gas accretion to regulate galaxy gas reservoirs has been supported by many cosmological simulations (see \citealt{Dekel2009,Voort2011a,Lagos2014,Nelson2015,Correa2018,Mitchell2020,Wright2020}). Observationally, however, the diffuse nature and weak kinematic signature of accreting IGM gas makes its direct detection an arduous pursuit (e.g. \citealt{Tumlinson2011,Sanchez2017}).

In light of these challenges, observational studies have largely used continuity arguments instead of direct detection to demonstrate the role smooth gas inflow (i.e. gas {\it not} accumulated via mergers) from the inter-galactic medium (IGM) in regulating galaxy properties (e.g. \citealt{Dave2012,Lilly2013}). For instance, it has been established that the consumption timescale of galactic gas is much shorter than a Hubble time, and that continued smooth gas accretion is required to sustain the observed cosmic star formation to $z=0$ (see \citealt{Kennicutt1983,Sancisi2008,Madau2014,Sanchez2017,Williams2020,Chowdhury2020}). 

Recently, a number of observational studies have presented direct detection of gas inflow in the CGM using metal-line absorption against stellar spectra, or neutral gas absorption against a fortuitously aligned QSO background source (e.g. \citealt{Rubin2012,Bouche2013,Fox2014,Bouche2016,Rubin2017,Turner2017,Bish2019,Roberts2019,Zabl2019,Herrera2020,Berta2020}). Even with these discoveries, the rate at which the gas accretes and its associated properties (such as its metallicity, temperature, and spatial distribution) have yet to be well-quantified on a statistical basis. and are associated with considerable uncertainties. Additionally, the metallicities that have been measured are relatively high, close to the metallicity of the galaxy the gas is being accreted onto -- suggesting that these detections correspond to recycling gas that had previously been ejected from the galaxy by stellar or Active Galactic Nuclei (AGN)-driven outflows (see \citealt{Marasco2012,Fraternali2015,Sanchez2017,Kacprzak2019}). 

Numerical simulations have highlighted the fact that gas recycling on both galaxy- and halo-scales only represents part of the picture of smooth gas inflow. Another key component is the gas on ``first-infall'' to haloes and galaxies from the cosmic web, with inflow aided by the gravitational pull provided by the collapsing regions around haloes. Cosmological simulations have found that gas on ``first-infall'' (that is, gas which has never been accreted onto a halo or galaxy previously) provides the majority of baryon growth to haloes at all epochs (with the exception of group- and cluster-sized haloes at $z\approx0$ in which mergers and recycling are dominant, see \citealt{Wright2020,Mitchell2020}). first-infall gas, having never been processed in a halo or galaxy environment, is often referred to as ``pristine'' or ``primordial'', based on the assumption that such gas has been seldom metal-enriched. One would expect such gas to be depleted in metal content relative to a ``recycled'' sample of accreting gas. This logic supports the {\it a priori} assumption made in many studies in the absence of kinematic data, in which lower metallicity gas is assumed to trace inflowing gas, while higher metallicity gas traces the outflowing component (e.g. \citealt{Kacprzak2012b,Kacprzak2015,Nielsen2020}). 

Smooth accreting gas is also often classified into a ``hot''-mode and a ``cold''-mode. In haloes with mass exceeding $\sim10^{12}M_{\odot}$, gas on infall towards over-densities is expected to be shock-heated -- an idea first explored in \citet{Rees1977,White1978,Binney1977}, and extended to the cold dark matter (CDM) paradigm in \citet{White1991}. Hot-mode accretion typically refers to gas that has been shock-heated and subsequently cools radiatively to join the central galaxy, while cold-mode accretion refers to the gas that accretes directly to the central galaxy in a filamentary manner, associated with the cosmic web. Numerical simulations have since demonstrated that hot-mode and cold-mode gas accretion can occur simultaneously, in a balance depending principally on halo mass \citep{Keres2005,Dekel2006,Ocvirk2008,Dekel2009,Voort2011a,Voort2012,Nelson2013,Correa2018a,Correa2018,Stern2020}.

 A combination of observations and simulations have shown that gas inflow has a measurable impact on the interstellar medium (ISM) of galaxies. Using SDSS data, \citet{Mannucci2010} and \citet{LaraLopez2010} independently found a fundamental plane (commonly referred to as the fundamental metallicity relation, FMR) between the properties of stellar mass, star-formation rate (SFR), and inter-stellar medium (ISM) metallicity in galaxies. Specifically, it was found that the scatter in the stellar mass-metallicity relation (MZR) at a fixed stellar mass correlates with SFR, with star-forming (quiescent) galaxies associated with low (high) metallicities. Pristine gas inflow provides a natural link between metallicity and SFR, with accretion encouraging star-formation while simultaneously ``diluting'' the ISM with enough low-Z gas to reduce its bulk metal fraction (e.g. \citealt{Kacprzak2016}). 
 
 This idea has been supported by simulations, which have found that gas accretion onto galaxies regulates ISM metallicities.  \citep{Collacchioni2019,DeLucia2020,vanLoon2021}. An indirect tracer of this regulation is via the gas content of galaxies. Simulations have shown that the scatter in the MZR is better correlated with the gas content of galaxies rather than their SFR (e.g. \citealt{Lagos2016,deRossi2017,DeLucia2020}. This is supported by observations, which find the atomic or molecular hydrogen content of galaxies to better describe the scatter in the MZR than SFRs (e.g. \citealt{Hughes2013,Bothwell2016,Brown2018}).
 
 Gas inflow also leaves a measurable imprint on the CGM surrounding galaxies. Observations have found a bimodality in the azimuthal angle of metal-line absorbers, with gas covering fractions enhanced by $20-30$\%  near the major and minor axes of galaxies (e.g. \citealt{Bordoloi2011,Kacprzak2012b,Kacprzak2015}). CGM gas detected on the major and minor axes of galaxies is proposed to be associated with co-planar,  inflows and feedback-driven outflows, respectively; a conjecture which has been reproduced in large-scale simulations (e.g. \citealt{Stewart2011,Voort2012,Shen2012,Peroux2020}). Assuming outflow winds are metal enriched relative to inflows, one would then expect to find an azimuthal dependence on the metallicity of CGM absorbers \citep{Lehner2016L,Peroux2020}. While a wide range of CGM metallicities have been measured (with a range of $>2$~dex, e.g. \citealt{Lehner2013,Prochaska2017,Zahedy2019}), the expected azimuthal dependence has not yet been confirmed with observations \citep{Peroux2016,Kacprzak2019,Pointon2019}. 
 
 Another notable accretion-regulated halo property is gas temperature, with the CGM of haloes nominally having both a hot and cold-phase. The hot coronal gas phase (at $\approx T_{\rm vir}$) originates from the virial shock-heating of gas accreting high-mass haloes ($M_{\rm halo}\gtrsim10^{12}M_{\odot}$; e.g. \citealt{Rees1977}). A cold-phase of the CGM at $\approx10^{4}$K has also been observed, but its origins are less clear (e.g. \citealt{Adelberger2003,Stocke2006,LehnerHowk2011,Prochaska2013,Zhu2014,Werk2014,Heckman2017,Zahedy2019}). Several origins of this cool CGM phase have been proposed, namely pristine IGM accretion (e.g. simulation-based findings in \citealt{Voort2012,Afruni2019,Afruni2020}), the condensation of hot halo gas (e.g. the empirical arguments of \citealt{Voit2018} and {\sc illustris-TNG} findings in \citealt{Nelson2020}), feedback-driven outflows (e.g. \citealt{Bouche2013,Borthakur2015,AnglesAlcazar2017,Oppenheimer2018,Hafen2020}), and the stripping of satellite galaxies in larger systems (e.g. \citealt{Hafen2020} using the {\sc fire-2} simulations). \citet{Afruni2019,Afruni2020}, using semi-analytic models and results from the COS-Halos and COS-GASS surveys, argue that star-formation driven outflows cannot account for the amount of cool gas in the CGM of observed haloes, pointing towards IGM accretion as the origin of this gas. Additionally, the radial variation of CGM properties was explored in \citet{Fielding2020} using a number of hydrodynamical simulations (as part of the {\sc smaug} project). They find that the properties of the outer-CGM (at $\gtrsim0.5R_{\rm 200,\ crit}$) are shaped by larger-scale processes, such as cosmological accretion, rather than galactic feedback which dominates the inner regions, $\lesssim0.5R_{\rm 200,\ crit}$. In any case, the wide range of observed metallicities and temperatures observed implies that the CGM is a diverse, multi-phase gas reservoir, making it an ideal laboratory to study the influence of cosmological inflows.
 
 Recent IFU-based studies of the CGM continue to improve our understanding of resolved CGM properties and kinematics (see \citealt{Schroetter2016,Nielsen2020}, and references therein). New ways of probing the CGM continue to be proposed and tested, for instance using fast radio bursts (FRBs; \citealt{Macquart2020}), and observations of Ly$\alpha$ and metal-lines in emission (see overview and predictions in \citealt{Lokhorst2019} and \cite{Augustin2019}, who use the \eagle\ simulations and dedicated zoom simulations respectively). Detecting CGM gas in emission is currently possible at high redshifts ($z\gtrsim2$) using ground-based telescopes --  e.g. the Very Large Telescope (VLT) or Keck -- in conjunction with methods of enhancing the signal (e.g. \citealt{Steidel2011,Wisotzki2018}). This is also possible at lower redshift using space-based telescopes, such as the Hubble Space Telescope (HST; e.g. \citealt{Hayes2016}). A promising future space-based endeavour is the LUVOIR (Large UV/Optical/Infrared Surveyor), slated to provide spectroscopic sensitivity in the UV enhanced by $30-100$ times compared to the HST/Cosmic Origins Spectrograph instrument \citep{Luviour2019}. With improvements in methodology and instrumentation, it may become possible to distinguish different phases of the CGM based on spatially resolved properties; for instance differentiating between recycled and pristine inflow based on metallicity measurements. 
 
 In this study we use the \eagle\ suite of hydrodynamical simulations to characterise the nature of gas accreting to haloes from the IGM. Specifically, we investigate its history, spatial characteristics, metallicity, density, and temperature in an effort to predict observational signatures of accreting IGM gas, and its influence on integrated galaxy and CGM properties. This paper is organised as follows: in \S \ref{sec:methods}, we introduce (i) the \eagle\ hydrodynamical simulation suite and the sub-grid models that are relevant to this study, (ii) \velociraptor\ and \treefrog: the phase-space structure finder we use to identify bound haloes and substructures (and its accompanying halo merger tree generator), and (iii) \chumm: the code we use to calculate and analyse accretion rates onto haloes in \eagle. \S \ref{sec:s3} explores the spatial characteristics of gas accreting to haloes over redshift; \S \ref{sec:s4} explores the chemical enrichment of this gas as a function of halo mass and its position in density-temperature phase space; and in \S \ref{sec:s5} we discuss the influence of gas accretion rates on central galaxy and CGM properties. To conclude, in \S \ref{sec:conclusion} we summarise our findings and discuss the implications of our results on semi-analytic models and observations, together with future scientific directions. 

\section{Methods}\label{sec:methods}
Here we briefly outline our methodology in calculating accretion rates to haloes and classifying inflow channels. For a full overview of our method (as well as temporal and resolution convergence tests), we refer the reader to Section 2 of \citet{Wright2020}.

\subsection{The \eagle\ simulations}\label{sec:methods:eagle}
\begin{table*}
\begin{center}
\include{Tables/Simulation-table}
\caption{Simulation parameters for the \eagle\ runs utilised in this paper \citep{Schaye2015,Crain2015}. $L_{\rm box}$ is the comoving box size of the simulation; $N_{\rm part}$ refers to the number of DM particles (and initial number of gas particles); $M_{\rm DM}$ and $M_{\rm gas}$ refer to the masses of DM and gas particles in the simulation respectively; SPH refers to the smoothed particle hydrodynamics scheme used; $\epsilon$ refers to the Plummer equivalent maximum gravitational smoothing length; $\Delta T_{\rm SN}$ and $\Delta T_{\rm AGN}$ are the heating temperatures adopted for stellar and AGN feedback; and $N_{\rm field\ halo}\ (z=0)$ describes the number of field haloes in each run at the final snapshot.\label{tab:s2:simulations}}
\end{center}
\end{table*}

 The \eagle\ (Evolution and Assembly of GaLaxies and their Environments) simulation suite \citep{Schaye2015,Crain2015} is a collection of cosmological hydrodynamical simulations that follow the evolution of galaxies and cosmological structure down to $z=0$.  The ANARCHY \citep{Schaller2015} set of revisions, designed to correct for ``classical'' smoothed particle hydrodynamics (SPH) issues, were implemented in the \gadget-3 tree-SPH code \citep{Springel2005} to perform the \eagle\ simulations over a variety of periodic volumes and resolutions. \eagle\ adopts the parameters of a ${\Lambda}$CDM universe from \citet{Planck2014}, with initial conditions outlined in \citet{Jenkins2013}. Sub-grid physics models are included for important processes that occur on scales below the resolution-scale of the simulation, including (i) radiative cooling and photoheating, (ii) star formation, (iii) stellar evolution and enrichment, (iv) stellar feedback, and (v) supermassive black hole (SMBH) growth and AGN feedback. Below, we provide a brief description of how these mechanisms are modelled in \eagle. 

Photo-heating and radiative cooling are applied based on the work of \citet{Wiersma2009}, including the influence of 11 elements: H, He, C, N, O, Ne, Mg, Si, S, Ca, and Fe \citep{Schaye2015}. The UV and X-ray background described by \citet{Haardt2001} is applied on each element individually. Since the \eagle\ simulations do not provide the resolution to model cold, interstellar gas, a density-dependent temperature floor (normalised to $T=8,000$~K at $n_{\rm H}=10^{-1}{\rm cm}^{-3}$) is imposed. To model star formation, a metallicity-dependent density threshold is set, above which star formation is locally permitted \citep{Schaye2015}. Gas particles are converted to star particles stochastically, with the star formation rate based on a tuned pressure law \citep{Schaye2008}, calibrated to the work of \citet{Kennicutt1998} at $z = 0$.

The stellar feedback sub-grid model in \eagle\ accounts for energy deposition into the ISM from radiation, stellar winds, and supernova explosions. This is implemented via a stochastic thermal energy injection to gas particles in the form of a fixed temperature change, at $\Delta T = 10^{7.5}$K - there being no explicit injection of kinetic energy . The average energy injection rate from young stars is given by $f_{\rm th}\times 8.73\times10^{15}$ erg $g^{-1}$ of stellar mass formed, assuming a \citet{Chabrier2003} simple stellar population and that $10^{51}$ erg is liberated per supernova event. $f_{\rm th}$ is set by local gas density and metallicity, ranging from $f_{\rm th,\ min}=0.3$ to $f_{\rm th,\ max}=3$ in the fiducial \eagle\ model. For a value of $f_{\rm th}=1$, the number of feedback events per particle is of order unity.

 SMBHs are seeded in \eagle\ when a DM halo exceeds a virial mass of $10^{10}\,\rm h^{-1} M_{\odot}$, with the seed SMBHs having an initial mass of $10^{5}\,\rm h^{-1} M_{\odot}$. Subsequently, SMBHs can grow via Eddington-limited-accretion \citep{Schaye2015}, as well as mergers with other SMBHs, according to work by \citet{Springel2005b}.  Similar to stellar feedback, AGN feedback in \eagle\ also involves the injection of thermal energy into particles surrounding the SMBH in the form of temperature boost of ${\Delta}T_{\rm BH}=10^{8}$K (in the reference physics  run; \citealt{Schaye2015}). The rate of energy injection from AGN feedback is determined using the SMBH accretion rate, and a fixed energy conversion efficiency, as in Equation \ref{eq:agnfb}:
 
 \begin{equation}
\frac{\Delta E}{\Delta t}=\epsilon_{\rm f}\epsilon_{\rm r}\ \dot{m}_{\rm accr}c^{2},
  \label{eq:agnfb}
\end{equation}

\noindent{where $\dot{m}_{\rm accr}$ is a modified Bondi-Hoyle accretion rate (see Equations 9,10 in \citealt{Schaye2015}), and $\epsilon_{\rm f}\epsilon_{\rm r}=0.015$.}

There are several free parameters in a number of the \eagle\ sub-grid modules, and these are calibrated to match $z\approx0$ observations of (i) the galaxy stellar mass function, (ii) the galaxy size-mass relation, and (iii) the galaxy BH mass - stellar mass relation. This calibration is performed separately for reference and recalibrated resolutions. In the main body of this paper, we make use of 3 standard resolution \eagle\ runs, all with varying feedback physics but identical starting mass resolution (first three rows of Table \ref{tab:s2:simulations}). In Appendix \ref{fig:apdx:res:fcov}, we make use of two simulation boxes: a run using a more rudimentary, \gadget-like SPH implementation (L50-OLDSPH) otherwise identical to L50-REF, and also the higher resolution recalibrated run (L25-RECAL), to investigate the resolution convergence of our results. For the remainder of the paper, we refer to each run by their identifier in the ``run-name'' columns for brevity. 

\subsection{Structure finding and halo trees with \velociraptor\ and \treefrog}\label{sec:s2:velociraptortreefrog}

We identify haloes and subhaloes in the \eagle\ runs using \velociraptor\ \citep{Elahi2011,Elahi2019a,Canas2019}, a 6D friends of friends (6D-FOF) structure finding algorithm. \velociraptor\ first uses a 3D-FOF algorithm \citep{Davis1985} to identify field haloes, and subsequently applies a 6D-FOF algorithm (including spatial and velocity information) in order to separate virialised structures \citep{Elahi2019a}. Once the 6D-FOF algorithm has been run over a 3D-FOF object, any nested density peaks will be identified as ``sub-haloes'' of the parent halo. To link haloes through time, we use the halo merger tree code \treefrog\ \citep{Elahi2019b}, developed to work on the outputs of \velociraptor. This code compares the particles in haloes across multiple snapshots by calculating a ``merit'' based on the fraction of particles that are shared by two (sub)haloes $i$ and $j$ at different times. 

\subsection{Accretion calculations with \chumm}\label{sec:s2:chumm}
\begin{table*}
\begin{center}
\include{Tables/Classification-table}
\caption{A summary of the decomposition of accreting particles into distinct inflow channels. The first-infall, pre-processed and merger channels are based on the \velociraptor-generated particle history classifications in \citet{Wright2020}, and the hot and cold- modes are based on post-accretion particle temperatures.\label{tab:s2:classification}}
\end{center}
\end{table*}

In order to calculate accretion rates onto haloes, we developed and used the code package \chumm\ (Code for Halo AccUMulation of Mass, available at \url{https://github.com/RJWright25/CHUMM}). Our method, which focuses on the build-up of matter on halo-scales, is outlined in detail in \citet{Wright2020}. Like \citet{Voort2017} and \citet{Correa2018}, we calculate accretion over the interval between adjacent \eagle\ snapshots ($29$ snapshots from $z=20$ to $z=0$), corresponding to a $\Delta t$ ranging between $\approx 250\ {\rm Myr}$ at minimum (at $z\approx4$), and $\approx 0.9\ {\rm Gyr}$ at maximum (at $z\approx0$). The varying timesteps in \eagle\ that we use to calculate accretion rates correspond to $0.5-1.0$ times a halo dynamical time, and insofar as accretion calculations, we argue that the standard snapshots offer adequate cadence. The sensitivity of our accretion calculations to the time interval was explored in Appendix C in \citet{Wright2020}. 

To identify haloes, \chumm\ either uses the outputs from \velociraptor\ and selects particle members of 6D-FOF objects, or \chumm\ can use a spherical overdensity (SO) criterion, where particles within a spherical region defined by $R_{\rm vir}$ are selected as constituting a halo. For the purposes of this work, we exclusively use the FOF-based classification of particles into haloes, meaning we make no assumption about the morphology of haloes. In this paper, we only consider accretion rates to field haloes (i.e, not subhaloes), which may or may not contain substructure. To calculate accretion rates onto haloes at a snap $n$, we identify accretion candidates as the particles that exist in the halo at snap $n$ as per the definition above, but did not exist in the halo at snap $n-1$. The summed mass of these candidate particles, normalised by $\Delta t = t_{n-1} - t_{n}$ (where $t_{n}$ represents the lookback time at snap $n$), constitutes the raw {\it gross} total accretion rate of the halo at snap $n$. Accretion rates are split by particle type, with the particle type categorised at the initial snap $n-1$, before undergoing any processing in the halo (such that gas particles at snap $n-1$ which were transformed to star particles by snap $n$ would be considered gas inflow, not stellar inflow). 

We subsequently categorise the nature of the inflow particles (their accretion ``channel'' or ``mode'') based on (i) their host at snap $n-1$, and (ii) their processing history. The particle's host at snap $n-1$ determines the origin of accretion as either from the field, or from another \velociraptor\ structure (the particles of each origin we refer to as  ``cosmological''/``smooth'' or ``merger''/``clumpy'' accretion particles respectively). 

For the cosmological/smooth accretion case, a particle is considered  ``pre-processed'' if it has existed in any halo (as defined by \velociraptor) up to and including snap $n-1$ (the initial snap), and {``unprocessed''/``first-infall''} otherwise. Commonly the term ``pristine'' is used to describe the accretion channel of metal-poor, unenriched particles, however we elect to use the term first-infall for our unprocessed channel to recognise that these particles are only ``unprocessed'' insofar as \velociraptor's ability to identify bound structures, ultimately limited by the finite mass resolution of the simulation. We can further decompose the ``pre-processed'' channel of cosmological accretion into a ``recycled'' and ``transfer'' component - for particles which were previously processed in a progenitor (main, non-main, or satellite) halo; and those that were previously processed in an unrelated halo respectively (as in \citealt{Wright2020}). For the purposes of this work, we choose to not decompose the pre-processed mode into its recycled and transfer components, due to the small contribution of the transfer component. We note that we only consider particle processing including and after snapshot $9$ ($z\approx4.5$) in the \eagle\ simulation due to data availability -- meaning that particles accreted prior to snap $9$, if subsequently ejected and re-accreted, would be considered first-infall. As such, we restrict our analysis to $z\lesssim3$, where there have been adequate snapshots since $z\approx4.5$ to confidently classify accretion in this way. 

For the purposes of this work, we also introduce a temperature-based classification of inflow channels for non-merger particles based on the suggested definition of hot- and cold-mode accretion in \citet{Correa2018a} and \citet{Correa2018}. These works show that in \eagle, due to the implementation of stellar and AGN feedback driven heating, using an instantaneous post-accretion particle temperatures is more appropriate than the previous maximum temperature of the particle ($T_{\rm max}$). Using a cutoff in $T_{\rm max}$ does not allow for post-feedback cooling, and could wrongly associate gas to the hot-mode of accretion when in reality the gas has cooled. We explore the distribution of $T_{\rm max}$ for accreting gas, and the necessity to allow for cooling, in Appendix \ref{sec:apdx:tmax}. We consider an accreting particle to be part of the hot-mode if it satisfies the temperature cut requirement in Equation \ref{eq:tcut}:

\begin{equation}
T_{\rm post-shock}\geq 10^{5.5} {\rm\ K},
\label{eq:tcut}
\end{equation}
\noindent{\rm where $T_{\rm post-shock}$ refers to the temperature of a particle post-accretion, at snap $n$. The hot- and cold- modes of accretion are then calculated as the summed masses of non-merger (smoothly accreted) particles meeting each criterion in Equations \ref{eq:cold} and \ref{eq:hot}:}

\begin{align}
\dot{M}_{\rm cold}&=\Sigma_{i}M_{i} (T_{\rm post-shock}<10^{5.5}\ {\rm K}) /\Delta t, {\rm\ and} \label{eq:cold} \\
\dot{M}_{\rm hot}&=\Sigma_{j}M_{j} (T_{\rm post-shock}\geq10^{5.5}\ {\rm K}) /\Delta t. \label{eq:hot} 
\end{align}

\noindent{Each of these inflow channel definitions, based either on particle history or particle temperature, are summarised in Table \ref{tab:s2:classification}.}

\subsection{Defining the spatial characteristics of inflow}
\label{sec:s2:spatial}
In order to classify the spatial nature of inflow in each of the aforementioned channŁls, we define a  ``covering fraction'', $f_{\rm cov}$, which quantifies the extent to which inflow is isotropic in nature. The covering fraction of a halo essentially corresponds the solid angle significantly occupied by accreting gas. More collimated/filamentary inflow corresponds to low covering fractions ($f_{\rm cov}\to0$), while more isotropic inflow are associated with high covering fractions ($f_{\rm cov}\to1$).

To calculate $f_{\rm cov}$, we use spherical coordinates to bin the space around each halo into $72$ bins in solid angle: $12$ in azimuth ($\phi$) and $6$ in elevation ($\theta$), all equally spaced. We impose no requirement on radial position of particles, and only consider their distribution in projected angular space about the halo center of mass. For each of the $72$ cells, $i$, and for each inflow channel, $j$, we determine the {\it expected} mass influx in each cell, $\langle \dot{m}_{i,\ j}\rangle$, if inflow were isotropic by scaling the total mass influx of that mode, $\dot{M}_{j}$, by the solid angle, $\Omega_{i}$, of each cell:
\begin{equation}
    \langle \dot{m}_{i,\ j}\rangle=\dot{M}_{j}\times \frac{\Omega_{i}}{4\pi}.
    \label{eq:expm}
\end{equation}

\noindent{We then classify a cell, $i$, ``occupied'' or ``covered'' by inflow particles of mode $j$ if the actual cell inflow rate $\dot{m}_{i,\ j}$ exceeds a minimum fraction, $f$, of the expected isotropic inflow rate $\langle \dot{m}_{i,\ j}\rangle$:}
\begin{equation}
    \dot{m}_{i,\ j}\geq f\times \langle \dot{m}_{i,\ j}\rangle,
    \label{eq:occupied}
\end{equation}
\noindent{where we select $f$ to be $0.1$. We remark that altering this $f$ value between $0.05-0.3$ does not qualitatively influence our results, and only changes the normalisation. For each halo, the covering factor of mode $j$ is then calculated as the solid angle weighted fraction of occupied cells, as per Equation \ref{eq:fcov}:}
\begin{equation}
    f_{\rm cov,\ j}=\frac{\Sigma_{i}\Omega_{i}({\rm occupied,\ }j)}{4\pi}.
    \label{eq:fcov}
\end{equation}
In order to ensure that the numerical value of $f_{\rm cov}$ is not driven by the number of inflow particles, we only calculate $f_{\rm cov}$ for a mode $j$ if the inflow to that halo exceeds $10^3$ particles (where then the minimum expected mass flux, $f\times \langle \dot{m}_{i,\ j}\rangle$, corresponds to at least $1$ accreted gas particle: at minimum $\approx 1/f=10$ in each cell, and $10\times72=720$ particles in total). This is the case for $\approx 95\%$ haloes for each of the non-merger accretion modes in the mass range $M_{\rm halo}\gtrsim10^{12}M_{\odot}$ at all redshifts considered. The particle flux requirement only significantly reduces the number of haloes we can calculate merger covering fractions for, where we could only use $\approx 50\%$ of the sample towards $z=0$ due to less frequent merger accretion events.



\section{The history, temperature, and spatial characteristics of accreting IGM gas}
\label{sec:s3}

\begin{figure*}
    \centering
    \includegraphics[width=1\textwidth]{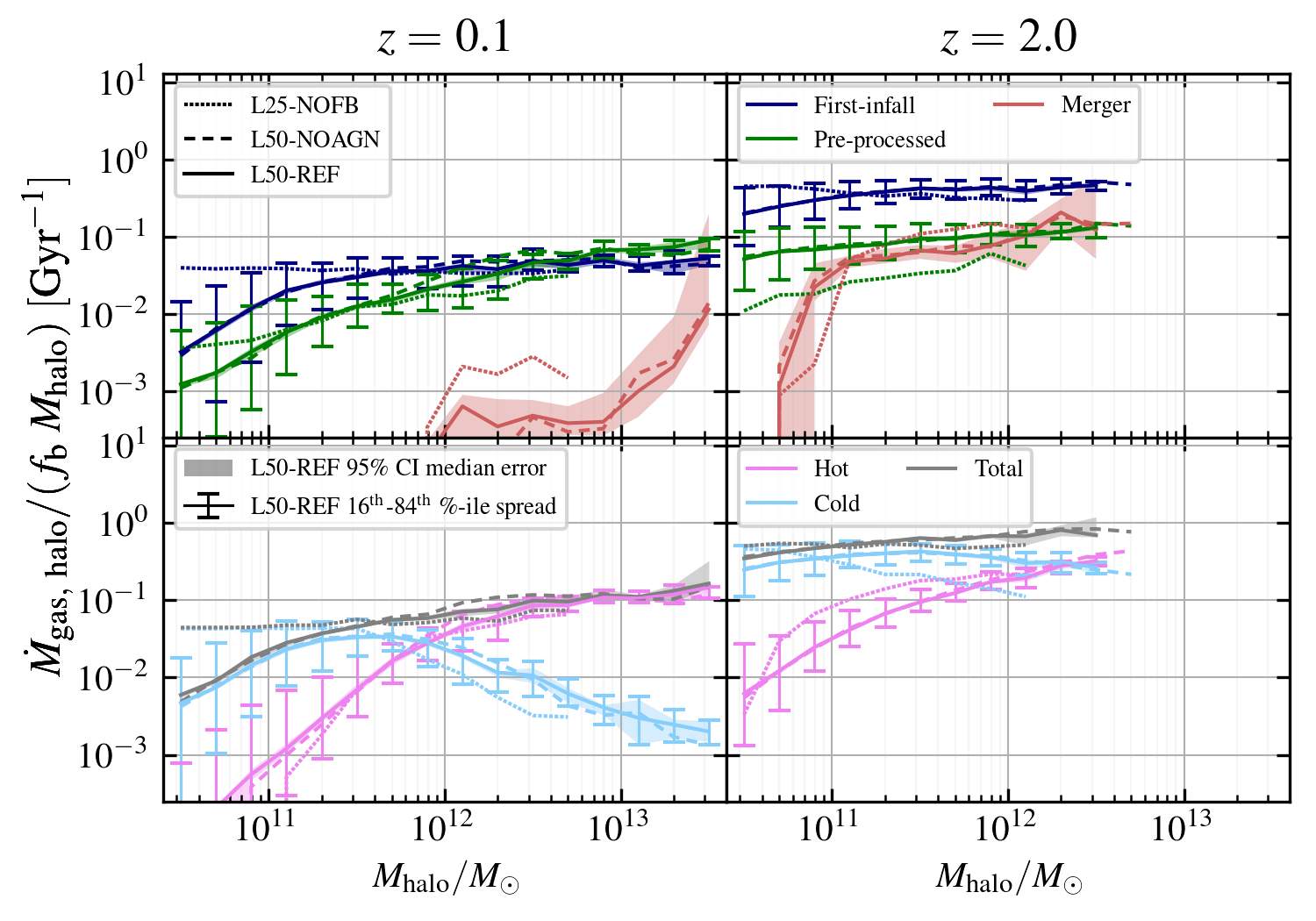}
    \caption{The median gas accretion efficiency, $\dot{M}_{\rm gas}/(f_{\rm b}M_{\rm halo})$, of each inflow mode in Table  \ref{tab:s2:classification}, as a function of halo mass at $z\approx0$ (left panels) and $z\approx2$ (right panels). The top panels include the history-based channel classifications (first-infall, pre-processed, and merger) and the bottom panels include the temperature based classifications (hot- and cold-mode accretion), together with total accretion rates. These accretion efficiencies are shown for the L50-REF run (solid lines), the L50-NOAGN run (dashed lines), and the L25-NOFB run (dotted lines). We also include the $16^{\rm th}-84^{\rm th}$ percentile range spread in gas accretion efficiency for L50-REF as error bars (excluding the merger-mode), and the bootstrap-generated $95\%$ confidence interval (CI) error on the median for L50-REF as shaded regions.}
    \label{fig:s3:channelefficiencies}
\end{figure*}

In this Section, we investigate the link between the temperature, history, and spatial characteristics of  gas accreting to haloes in \eagle. We focus on the halo mass range $M_{\rm halo}\gtrsim10^{11}M_{\odot}$, where, on average, inflow corresponds to a flux of more than $100$ gas particles over the designated time interval between each snapshot. We quantify and analyse gas inflow to these haloes from $z\approx3$ to $z\approx0$. Wherever we take bins in halo and stellar mass, unless otherwise stated, they are spaced in increments of $0.2$~dex. 

\subsection{Revisiting the breakdown of gas accretion rates}
\label{sec:s3:channels}

We start by revisiting the gas accretion rates originally presented in \citet{Wright2020}, in which accreting particles were broken down into being either first-infall, recycled, transfer, or merger-based in origin. As described in Section \ref{sec:s2:chumm} and Table \ref{tab:s2:classification}, for the purposes of this work, we combine the recycled and transfer components into one ``pre-processed'' mode, and add in a separate temperature-based inflow classification, originally conceived in \citet{Correa2018a}. In Figure \ref{fig:s3:channelefficiencies}, we illustrate the median gas accretion efficiencies, $\dot{M}_{\rm gas}/(f_{\rm b}\ M_{\rm halo})$, for each accretion channel as a function of halo mass for (i) the original history-based classifications (top panels), and (ii) the new temperature-based classifications (bottom panels) at $z\approx0$ (left panels) and $z\approx2$ (right panels). We use the value of $f_{\rm b}$ implied by the results of \citet{Planck2014}, at $\Omega_{\rm b}/\Omega_{\rm m}=0.157$. 

The top panels illustrate the trends seen in Figures 6 and 10 of \citet{Wright2020}, instead showing the normalisation of total accretion rates (as opposed to the fractional contribution of each mode). We comment on the physical accuracy of our infall channel categorisations in Appendix \ref{sec:apdx:tmax}. In this Figure and throughout the rest of the paper, solid lines represent results from the L50-REF reference physics run, dashed lines represent results from the L50-NOAGN run, and the dotted lines correspond to results from the L25-NOFB run. Additionally, the $16^{\rm th}-84^{\rm th}$ ($1\sigma$) percentile range is shown with errorbars, and the bootstrap-generated $95\%$ confidence interval error on the median (from 100 resamples using a randomly selected half of the appropriate population) is illustrated as a shaded region for the L50-REF reference run exclusively. 

\begin{figure*}
    \centering
    \includegraphics[width=1\textwidth]{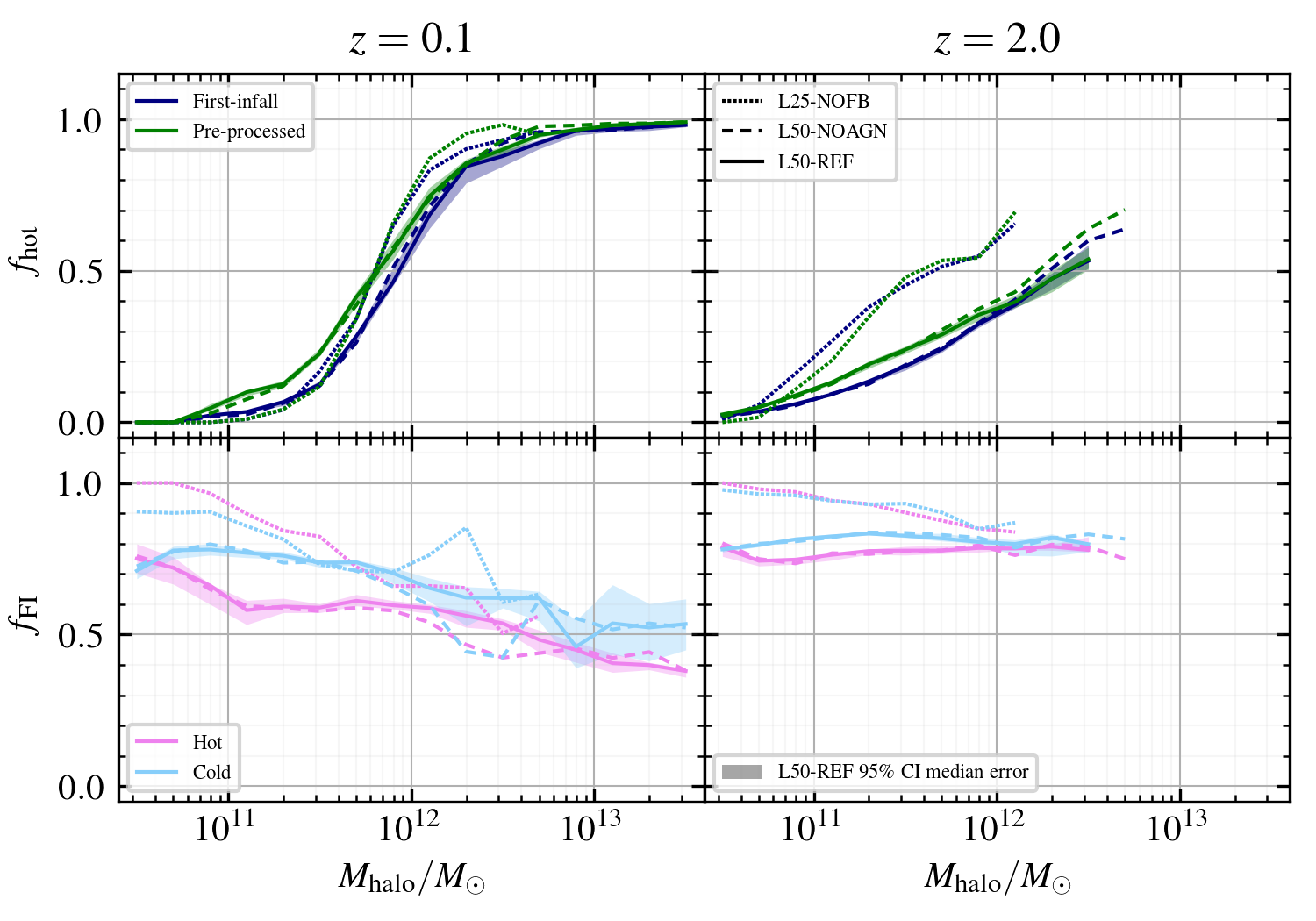}
    \caption{Top panels: The median hot fraction (defined by the temperature cut in Equation \ref{eq:tcut}) of inflow gas for the first-infall and pre-processed modes as a function of halo mass at $z\approx0$ (left) and $z\approx2$ (right). Bottom panels: the median fraction of mass delivered in the form of unprocessed gas for hot- and cold- modes as a function of halo mass, for the same redshifts. An $f_{\rm hot}$ value of $0.5$ for the first-infall mode, for instance, would mean that half of the gas accreted via the first-infall mode was ``hot''; while a $f_{\rm FI}$ value of $0.5$ for the hot-mode would mean that half of the hot inflow gas was accreting to a halo for the first time in its history. Each of these fractions are shown for the L50-REF run (solid lines), the L50-NOAGN run (dashed lines), and the L25-NOFB run (dotted lines). We also include the bootstrap-generated $95\%$ confidence interval  error on the median for L50-REF as shaded regions.  }
    \label{fig:s3:fhot}
\end{figure*}

Concentrating first on the L50-REF reference physics run (solid lines), at $z\approx0$ we see that first-infall accretion dominates for $M_{\rm halo}\lesssim10^{12.5}M_{\odot}$, and pre-processed accretion dominates above this transition mass. The contribution of mergers to mass growth is ubiquitously small, but increases with halo mass to be within $1$~dex of the first-infall and pre-processed modes at $M_{\rm halo}\gtrsim10^{13}M_{\odot}$. If we compare this breakdown to that between the hot and cold- modes of accretion in the bottom left panel, we see qualitative similarity between cold and first-infall channels, as well as the hot and pre-processed channel - in that cold accretion dominates for $M_{\rm halo}\lesssim10^{12}M_{\odot}$, and above this transition mass, hot-mode accretion takes over. While we find these similarities between the first-infall (pre-processed) and cold (hot) modes of accretion, we are not claiming that these causally or physically linked: rather, we argue that the similarities are driven by a set of physical processes defined by similar halo mass transition scales. In the case of the breakdown between first-infall and pre-processed accretion, we show in \citet{Wright2020} that the pre-processed mode increases in flux with halo mass as haloes become massive enough to attract previously ejected particles (either from one of the halo's progenitors, or an unrelated halo). Conversely, in the hot/cold-mode case, the contribution of the hot-mode increases with halo mass as haloes become massive enough to efficiently shock heat the accreting gas (e.g. \citealt{Katz2003,Keres2005,Keres2009a,Ocvirk2008,Voort2011a}). Note that we  explore the relationship between the temperature and history-based inflow classifications in Figure \ref{fig:s3:fhot}, which we discuss further below.

At $z\approx 2$, we see that first-infall accretion channel exceeds pre-processed accretion for all halo masses, with a roughly constant offset of $\approx 0.6$~dex. Merger-based accretion efficiency again increases with halo mass, and roughly tracks the pre-processed channel above halo masses of $\approx10^{11.5}M_{\odot}$. We also find that cold-mode accretion dominates over hot-mode accretion for the full halo mass range, with the hot accretion channel increasing with halo mass to nearly meet the relatively constant cold-mode efficiency at $M_{\rm halo}\approx10^{12.5}M_{\odot}$. The rise in contribution of hot-mode accretion with halo mass is unsurprising, given the increased importance of virial shock-heating \citep{Binney1977,Rees1977,Birnboim2003,Dekel2006,Keres2009a}, however \citet{Correa2018a} show in \eagle\ that other physical processes are also required to explain gas accreting above halo virial temperatures. 

Focusing on the influence of sub-grid physics, we note an increase in $z\approx0$ total accretion efficiencies in the L25-NOFB run compared to L50-REF and L50-NOAGN at halo masses below $10^{11}M_{\odot}$ - a direct consequence of the lack of stellar feedback, as discussed in \citet{Wright2020}. At $z\approx2$, we find that while the first-infall mode in the L25-NOFB run is consistent with L50-REF, the pre-processed mode appears to be suppressed due to the lack of mechanisms to eject particles from haloes (for subsequent pre-processed re-accretion). At $z\approx2$, we note that hot-mode accretion in the L25-NOFB run is strongly enhanced relative to L50-REF, which we explore further in the discussion of Figure \ref{fig:s3:fhot}. We note that the influence of AGN feedback on accretion modes is minimal, with only a slight increase in total accretion rates for haloes between $10^{12}M_{\odot}$ and $10^{12.5}M_{\odot}$ at $z\approx0$ in L50-NOAGN compared to L50-REF, where we find that hot-mode inflow is slightly suppressed when AGN feedback is included. We explore this suppression further in the context of inflow covering fractions in \S \ref{sec:s3:spatial}. 

To investigate the link between the history-based and temperature-based classification of accretion modes, in the top panels of Figure \ref{fig:s3:fhot} we show the fraction of mass that accreted ``hot'' (see Equation \ref{eq:hot}) as a function of halo mass, for accreting gas identified as first-infall and pre-processed. $f_{\rm hot}$ fractions for both modes increase with halo mass at both $z\approx0$ and $z\approx2$, in a similar fashion to total hot accretion fractions presented in \citet{Correa2018}. Focusing on reference physics (solid lines), we find that the pre-processed mode is systematically ``hotter'' than the first-infall mode, with an offset in median hot fraction of $10\%$ for haloes below $M_{\rm halo}\approx10^{12}M_{\odot}$ at both $z\approx0$ and $z\approx2$. We remark that the bootstrapped error on the respective medians do not overlap below $M_{\rm halo}\approx 10^{12}M_{\odot}$ at each redshift, meaning we consider this to be a significant and real difference in the temperature of first-infall and pre-processed accreting gas. 

The bottom panels of Figure \ref{fig:s3:fhot} show the converse of the upper panels: instead of the fraction of particles accreting hot for each history-based inflow classification, we show the fraction of particles accreting for the first time ($f_{\rm first\ infall},{\rm\ or}\ f_{\rm FI}$), broken down into gas accreting via the hot and cold channels of inflow. 

At $z\approx0$, we see a trend for $f_{\rm FI}$ for both hot- and cold- modes to decrease with halo mass from $\approx75\%$ at $M_{\rm halo}\approx10^{10.5}M_{\odot}$ to $\approx50\%$ at $M_{\rm halo}\approx10^{13.5}M_{\odot}$ (in line with the global shift towards recycled and transferred baryonic accretion found in \citealt{Wright2020}, Figure 6). Embedded in this global trend, we see a significant disparity between the hot- and cold-mode unprocessed fractions for haloes in the mass range $10^{11}M_{\odot}\lesssim M_{\rm halo}\lesssim10^{12}M_{\odot}$ - with $f_{\rm FI}$ of hot-accreting gas being $\approx20\%$ lower than cold-accreting gas. In this halo mass range, cold-accreting gas is significantly more likely to be on first-infall than to have been previously processed. Above this halo mass range, $f_{\rm FI}$ is similar for the hot- and cold- modes, though we remind the reader that the cold-mode is also heavily suppressed in this regime at $z\approx0$. 

At $z\approx2$, the differences between hot- and cold-mode $f_{\rm FI}$ are qualitatively similar to those found $z\approx0$ using L50-REF. Values of $f_{\rm FI}$  remain at $\approx70\%$ for the full halo mass range (slightly higher than at $z\approx0$). There is still a significant systematic difference between cold and hot-mode median $f_{\rm FI}$ values in the mass range $10^{10.5}M_{\odot}\lesssim M_{\rm halo}\lesssim10^{11.5}M_{\odot}$, with cold-accreted particles $\approx10\%$ more likely to be on first-infall compared to being pre-processed. The spread in $f_{\rm FI}$ is lower for both hot- and cold- modes at $z\approx2$, however there is still overlap between the percentile ranges for the full halo mass range. 

Shifting focus away from the L50-REF run, we can use the L50-NOAGN and L25-NOFB curves in Figure \ref{fig:s3:fhot} to investigate the role of feedback in altering hot-accretion fractions and first-infall fractions. We remind the reader that in Figure \ref{fig:s3:channelefficiencies}, we found enhanced hot accretion in the L25-NOFB run relative to reference physics. In the top panels of Figure \ref{fig:s3:fhot}, we show that the median hot fraction of both the first-infall and pre-processed modes in the L25-NOFB run is significantly increased, particularly at $z\approx2$ where the hot-mode is $\approx20\%$ more prominent than in L50-REF for $M_{\rm halo}\gtrsim10^{11.5}M_{\odot}$, with raw total accretion rates otherwise very similar. 

Upon investigation, we attribute the enhanced hot fraction of accretion in L25-NOFB to two factors. Firstly, we find that the baryon fraction of haloes in the L25-NOFB run are larger than in L50-REF and L50-NOAGN run, and that this baryonic component is less radially extended than L50-NOAGN/REF haloes at a given mass. We consequently argue that the compact, gas-rich halo environment in L25-NOFB facilitates efficient shock heating of accreting gas. Secondly, we find that that the non-merger component of accreting to haloes in the L25-NOFB run is heavily metal-depleted relative to reference physics (see Figure \ref{fig:s4:massZ}, top panels). The low metallicity of accreting gas extends cooling times after virial heating - meaning that less accreting gas is  able to cool below $10^{5.5}$~K shortly after accretion, as would be required to be classified as ``cold''. 

The lack of metal-enriched accreting gas in L25-NOFB (compared to L50-REF) is a likely consequence of the lack of stellar-feedback driven outflows, which can act to enrich the CGM surrounding galaxies and haloes. With these outflows, accreting gas which has been previously processed (for instance, gas which has been ejected from a halo and is subsequently re-accreting) is more likely to have been enriched, and gas on first-infall is more likely to have been enriched when in contact with the outskirts of the halo environment prior to accretion. We explore these ideas further in \S \ref{sec:s4}. 

We find that AGN feedback plays a sub-dominant role in regulating hot accretion fractions at $z\approx0$, however at $z\approx2$, we note that both pre-processed and first-infall hot fractions in L50-NOAGN are slightly increased relative to L50-REF above $M_{\rm halo}\approx10^{12}M_{\odot}$. At maximum, we see a difference of $\approx10\%$ in medians at $M_{\rm halo}\approx10^{12.5}M_{\odot}$. This could be a result of slightly higher halo baryon fractions and a more compact baryonic profiles in the L50-NOAGN run at this halo mass, facilitating more efficient shock-heating of the accreting gas. The increase accretion hot fractions in L50-NOAGN are specifically a result of increase hot-mode accretion in relation to the cold-mode, which remains fairly similar between L50-NOAGN and L50-REF. This suggests that AGN feedback preferentially suppresses hot-mode accretion as opposed to the cold-mode, which we discuss further in \S \ref{sec:s3:spatial}.

\begin{figure*}
    \centering
    \includegraphics[width=0.9\textwidth]{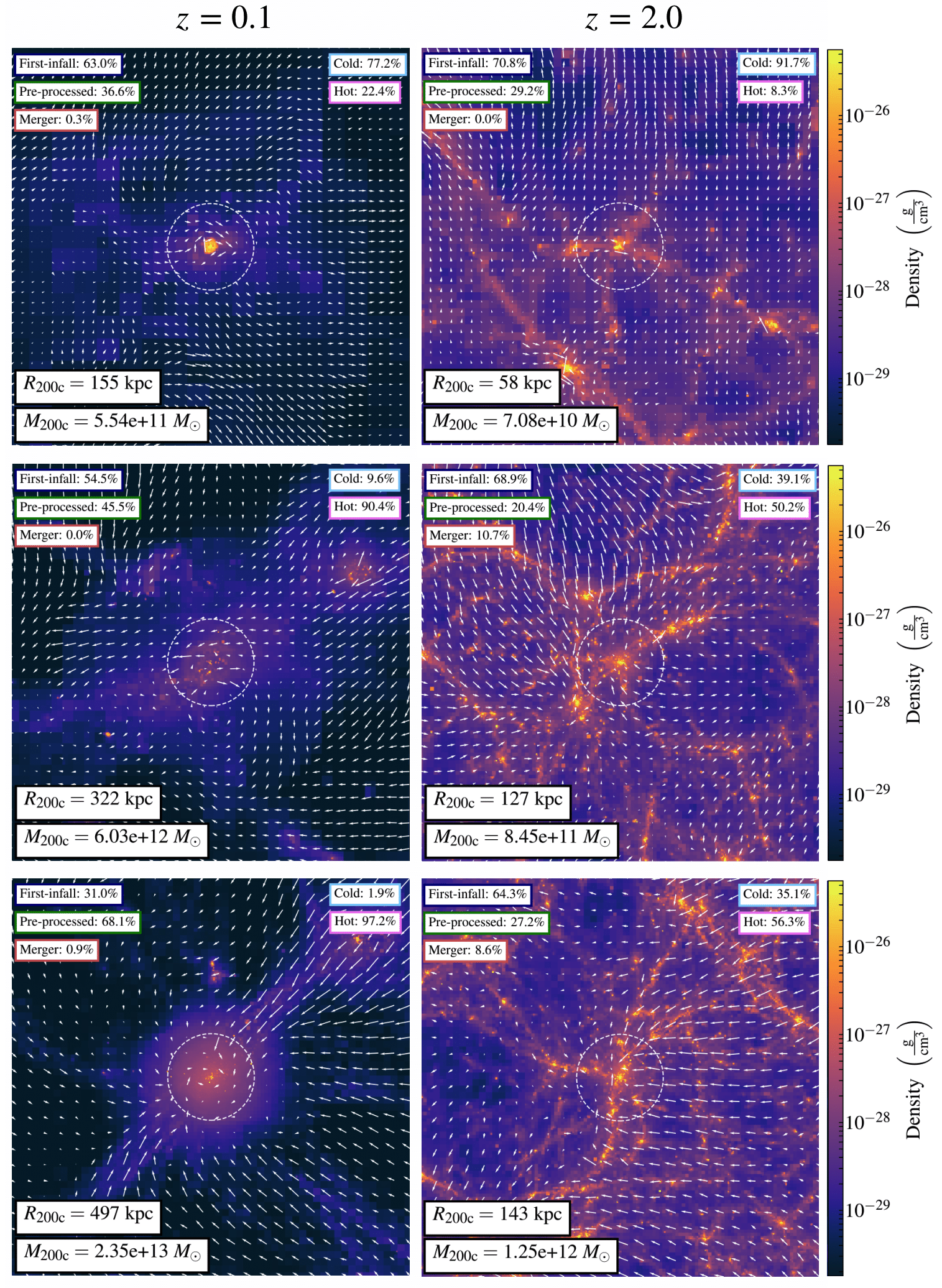}
    \caption{A visualisation of the gas surrounding 3 example haloes at $z\approx0$ (left panels) and their progenitors at $z\approx2$ (right panels). We include a halo at $\approx10^{11.5}M_{\odot}$ (``dwarf-mass'', top panels), a halo at $\approx10^{12.5}M_{\odot}$ (``Milky Way-mass'', middle panels), and a halo at $\approx10^{13.5}M_{\odot}$ (``group-mass'', bottom panels). Gas is coloured by log-scaled density, and  halo-centric averaged velocity vectors are overlayed with white arrows. Additionally, the virial sphere for each halo is circled in white, and the breakdown of accreting gas (which had entered the halo at the previous snap) into its constituent channels is quantified and quoted as a percentage of total inflow mass. Visualisations were produced with the {\it yt} package \citep{Turk2011}.}  
    \label{fig:s3:gasvisualisation}
\end{figure*}

\subsection{Spatial distribution of accreting gas}
\label{sec:s3:spatial}
Using the accretion classifications presented in Table \ref{tab:s2:classification} and statistically measured in Figure \ref{fig:s3:channelefficiencies}, we now investigate and quantify the spatial characteristics of gas accreting to haloes in \eagle\ based on the parameters in Section \ref{sec:s2:spatial}.

We visualise the gas accreting to $3$ example haloes at of different mass at $z\approx0$, and their progenitors at $z\approx2$, in Figure \ref{fig:s3:gasvisualisation}. We choose these $3$ haloes to demonstrate the nature of accreting gas for (i) dwarf-mass haloes: $M_{\rm halo}(z=0)\approx10^{11.5}M_{\odot}$, (ii) Milky Way (MW)-mass haloes: $M_{\rm halo} (z=0)\approx10^{12.5}M_{\odot}$, and  (iii) group-mass haloes: $M_{\rm halo} (z=0)\approx10^{13.5}M_{\odot}$. 

Gas accreting to the  $10^{11.5}M_{\odot}$ halo at $z\approx0$ (top left panel) is preferentially in the form of cold-mode inflow ($\approx77\%$) compared to the gas accreting to more massive haloes. The inflow is dominated by particles on first-infall to a halo ($\approx 63\%$) compared to the pre-processed mode ($\approx 37\%$), with negligible contribution by mergers. While the cold-mode of accretion dominates over the hot-mode, the accreting matter does not appear filamentary in structure. This is not the case for the halo's progenitor at $z\approx2$ (top right panel) which appears to be accumulating mass from a number of filaments. At $z\approx2$, we find that the halo has a similar breakdown between first-infall and pre-processed inflow ($\approx 71\%$ compared to $\approx 29\%$).  At $z\approx0$, rotation about the central galaxy is clear in the velocity fields, however gas velocity vectors at the virial sphere do not isotropically favour accretion. We observe a similar picture for the $z\approx2$ progenitor, with gas flows at the virial radius not necessarily favouring infall. 

Unlike the dwarf-mass halo, gas accreting to the $10^{12.5}M_{\odot}$ halo at $z\approx0$ (middle left panel) is $\approx90\%$ hot in nature, with only $10\%$ being accreted via the cold-mode. There is no merger contribution to the mass growth of the halo at this snapshot, with accreting gas roughly evenly split between gas on first-infall to a halo ($\approx 55\%$) and pre-processed gas ($\approx 45\%$). Compared to the $z\approx0$ descendent, the progenitor at $z\approx2$ sees more first-infall accretion ($69\%$) and cold-mode accretion ($39\%$) together with a significant merger contribution of $\approx11\%$.  Considering the $z\approx0$ velocity field, we observe a combination of inflow and outflow at the virial radius, with inflow particularly prominent in the top right quadrant. In this direction, it appears the halo is being fed by a bridge of gas connecting the halo to an approaching halo of similar mass. Unlike the $z\approx0$ descendent, we find that the velocity field around the $z\approx2$ halo mostly favours infall at the virial radius. 

Gas surrounding the  $10^{13.5}M_{\odot}$ halo at $z\approx0$ (bottom left panel) forms a hot, near-spherical atmosphere near $R_{200}$, fed by a number of filaments. The vast majority of recently accreted gas is hot in nature, with $\approx 97\%$ having been heated upon entering the halo. The non-merger component is dominated by pre-processed gas ($\approx68\%$) as opposed to gas on first-infall ($\approx 33\%$). Gas flow at the virial radius very clearly favours accretion to the halo, particularly along the surrounding filaments in the bottom left and top right quadrants. At $z\approx2$ (bottom right panel), the picture for the group's progenitor is quite different, there being a much more defined filamentary structure to the halo and surrounding regions, as opposed to a shock-heated hot halo. The recently accreted material is split between $\approx35\%$ cold-mode and $\approx56\%$ hot-mode, the remaining $\approx9\%$ originating from mergers. The non-merger history-based breakdown of accretion is dominated by first-infall ($\approx 64\%$) gas compared to particles which had been pre-processed ($\approx 27\%$). The gas velocity field at the virial radius mostly favours accretion to the halo, with the exception of the bottom left sector where gas is being attracted by another large structure. 

In general, we see that inflow is noticeably more filamentary and cold in nature for $z\approx2$ progenitors compared to their $z\approx0$ descendents for each of the example haloes, spanning 2 orders of magnitude in mass. Additionally, in the $z\approx0$ selected MW-mass and group-mass haloes, we find that merger-based gas accretion plays a more important role for their $z\approx2$ progenitors - consistent with a hierarchical growth scenario. At fixed redshift, we see that hot-mode accretion dominates for higher mass haloes compared to cold-mode accretion, which dominates gas accretion in the dwarf halo regime. These example observations agree with the trends in gas accretion efficiency shown in Figure \ref{fig:s3:channelefficiencies}. The qualitative trends we observe with the apparent filamentary structure of accreting gas can be quantified by the use of the spherical covering fraction introduced in Equation \ref{eq:fcov}, and demonstrated below in Figure \ref{fig:s3:fcov}.

\begin{figure}
\centering
    \includegraphics[width=1\columnwidth]{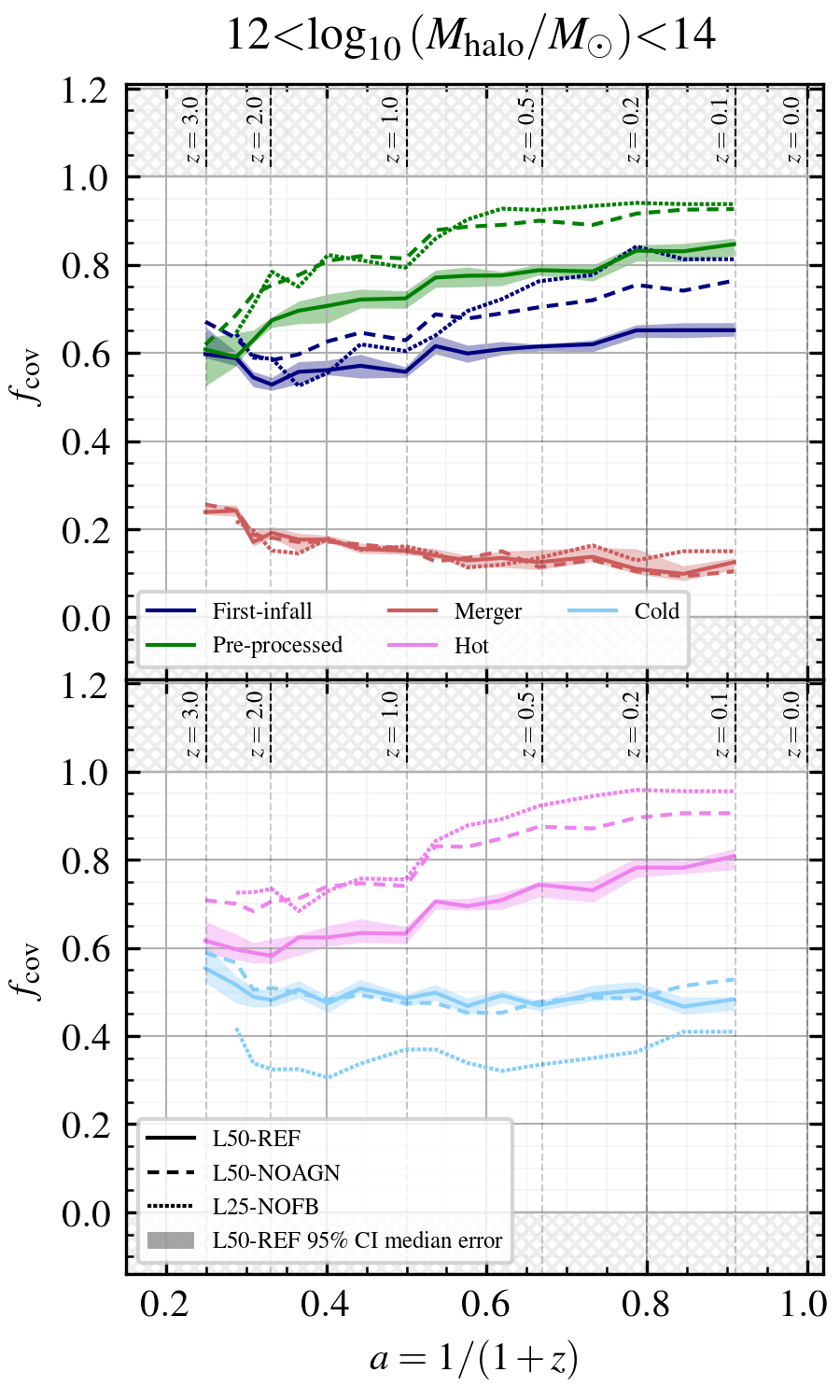}
    \caption{The median covering fraction, $f_{\rm cov}$ (as defined in Equation \ref{eq:fcov}), of accreting gas in haloes with $10^{12}M_{\odot}<M_{\rm halo}<10^{14}M_{\odot}$ as a function of redshift. We show $f_{\rm cov}$ individually for each accretion mode in Table \ref{tab:s2:classification} (with history-based classification in the top panel, and temperature based classification in the bottom panel), requiring at least 1000 particles accreted for each mode for a halo to be analysed). $f_{\rm cov}$ is shown for the L50-REF run (solid lines), the L50-NOAGN run (dashed lines), and the L25-NOFB run (dotted lines). We also include the bootstrap-generated $95\%$ confidence interval  error on the median for L50-REF as shaded regions. }
    \label{fig:s3:fcov}
\end{figure}

Using the definition of spherical inflow covering fraction, $f_{\rm cov}$, introduced in Equation \ref{eq:fcov}), we illustrate the redshift evolution of the spatial distribution of halo-scale accreting matter for each accretion mode in Figure \ref{fig:s3:fcov}. To ensure $f_{\rm cov}$ is not sensitive to the number of particles accreting (and is exclusively a reflection of the spatial {\it distribution} of accreting matter), we only include haloes at each snap that are (i) in the mass range $10^{12}M_{\odot}<M_{\rm halo}<10^{14}M_{\odot}$, and (ii) have accreted at least $10^{3}$ gas particles since the previous snapshot. We impose the latter for each of the accretion modes individually, meaning the halo sample for each curve is not identical. Within the imposed mass range, an accretion flux of $\geq10^{3}$ gas particles occurs in at least $\approx 95\%$ of haloes for all non-merger accretion modes (at all snapshots). The particle flux requirement only significantly reduces the number of haloes we can calculate merger covering fractions for, where we could only use $\approx 50\%$ of the sample towards $z=0$ due to infrequent merger accretion events.

First concentrating on the L50-REF run and the history-based accretion channels (top panel of Figure \ref{fig:s3:fcov}), we see that the covering fraction of pre-processed and first-infall gas modestly increases towards $z=0$. This indicates a shift from more filamentary, collimated inflow towards a more isotropic distribution, consistent with the qualitative picture painted in Figure \ref{fig:s3:gasvisualisation}. The first-infall and pre-processed modes possess a median covering fraction of at least $0.5$ for the full redshift range, but the pre-processed mode has a consistently larger $f_{\rm cov}$ parameter compared to first-infall accretion by $\approx 20\%$ over the majority of the redshift range. The median values prove to be significantly different based on the bootstrap-generated error ranges, indicating that particles on first-infall to a halo are significantly more likely to come in the form of more collimated inflow compared to particles which had been previously accreted and ejected by a halo. Qualitatively, this aligns with the scenario of first-infall particles originating from cosmic web filaments. The covering fraction of first-infall accretion reaches a maximum at $z\approx0$ of $\approx 70\%$ compared to the $\approx 90\%$ of the pre-processed mode. Unsurprisingly, the covering fraction of inflow from mergers is comparatively very low ($\approx 20\%$) as a result of the obvious unidirectional nature of such accretion events. 

Looking instead at the covering fraction of temperature-based accretion channels in the bottom panel of Figure \ref{fig:s3:fcov}, we can immediately note the offset between hot-mode accretion and cold-mode accretion - similar to the disparity between the pre-processed and first-infall $f_{\rm cov}$ values in the top panel. Unlike the first-infall mode, we don't find any significant evolution of cold-mode covering fractions with redshift, with cold-mode $f_{\rm cov}$ remaining at a roughly constant value of $\approx 50\%$, compared to the $>60\%$ found for the first-infall mode. Similar to the pre-processed mode, the hot-mode of accretion reaches a maximum $f_{\rm cov}$ value of $\approx 90\%$ at $z\approx0$. The evolution of $f_{\rm cov}$ for the cold and hot-modes is not necessarily expected. The qualitative picture that has been discussed in the literature tends to connect the hot-mode with isotropic accretion, with correspondingly large covering fractions, and cold-mode with highly collimated accretion, with much smaller covering fractions. We show here that the picture is more complicated than this, and in fact the covering fraction of cold-mode is quite large (albeit always smaller than that of the hot-mode) across the whole redshift range studied. hot-mode accretion, on the other hand, shows significant redshift evolution of its covering fraction, and at $z\approx3.0$ appears to be as filamentary as the cold-mode. 

Interestingly, we note that sub-grid physics significantly influences the spatial distribution of accreting gas in this halo mass range. In the absence of AGN feedback (L50-NOAGN), we see a significant enhancement (by $\approx 10-15\%$) of the first-infall, pre-processed, and hot-mode inflow covering fraction, with L50-NOAGN median $f_{\rm cov}$ values sitting near the $84^{\rm th}$ percentile in $f_{\rm cov}$ values from L50-REF across the full redshift range for these modes. In \citet{Wright2020}, we show that AGN activity can modulate accretion rates in this mass regime by $\approx30\%$, which appears to also reflect in the covering fraction of accreting matter. This conclusion agrees well with the findings of \citet{Nelson2015} using {\sc Illustris}-like physics, who find an enhancement in halo-scale spherical ``covering fraction'' in the absence of stellar and AGN feedback, particularly at high redshift. This suggests that AGN driven outflows have a real and quantifiable interaction with inflowing matter, acting to reduce the vacant angular fraction where gas is able to radially infall, even at the halo-scale. The preferential suppression of hot accretion by AGN (as discussed in relation to Figure \ref{fig:s3:channelefficiencies}) would be consistent with a picture where AGN-driven outflows can disturb the more isotropic hot-mode of inflow, but have less influence on the more collimated, cold-mode of inflow. 

In terms of $f_{\rm cov}$, the behaviour of the L25-NOFB run is similar to the L50-NOAGN run for the pre-processed, first-infall and hot channels of accretion. In contrast, focusing on the cold-mode, we note that covering fractions of cold-mode accreting matter decrease by $\approx 20\%$ in the absence of stellar feedback for the full redshift range analysed. In \eagle, \citet{Correa2018a} show that stellar feedback has a considerable influence on the amount of hot gas in the halo. They show for haloes at $M_{\rm halo}\approx10^{12}M_{\odot}$ that a doubling of stellar feedback strength leads to a increase in the gas mass fraction by a factor of $1.3$, and that a halving of stellar feedback strength reduces the gas mass fraction by a factor $2.5$. We suggest that the presence of a hot gaseous corona maintained by stellar feedback causes cold inflow gas to become less collimated as it approaches a halo, thereby increasing the covering fraction, $f_{\rm cov}$, of gas accreting via this mode in L50-NOAGN and L50-REF compared to L25-NOFB. This does not influence the already heated and less collimated hot-mode.

We remark that in Appendix \ref{sec:appendix:fcov}, we check the ``weak convergence'' (see \citealt{Schaye2015} for explanation) of our covering fractions, and find that $f_{\rm cov}$ values from the L25-RECAL run are largely consistent (within uncertainty) with L50-REF covering fractions over cosmic time. We also investigate the influence of SPH scheme on our $f_{\rm cov}$ calculations, and find that using an older, \gadget-like SPH implementation (without the improved Pressure-SPH scheme) produces systematically lower accretion covering fractions (see Figure \ref{fig:apdx:res:fcov}).


\section{The metal enrichment of accreting gas}
\label{sec:s4}

In this Section, we investigate the metallicity of gas accreting to haloes in \eagle, in particular the level of enrichment for gas accreting via different inflow channels (as described in Table \ref{tab:s2:classification}). For our calculations, we use individual particle metallicities that have not been smoothed over the SPH kernel. While the smoothed metallicity values are used in cooling calculations, this smoothing could contaminate otherwise pristine particles, which we wish to capture in our analysis.

We also remark that when we refer to the ``integrated'' metallicity of the matter accreting to a halo, we calculate the total mass accreted in metals normalised by the total mass accreted in gas: $Z_{\rm int}={\dot{M}_{\rm metals}/\dot{M}_{\rm gas}}=\Sigma_{i}(Z_{i}\times M_{i})/\Sigma_{i}(M_{i})$, summing over all accreted particles $i$. This measurement is typically skewed towards higher values of $\log_{10}(Z/Z_{\odot})$ compared to the median metallicity of individual accreting particles. 
\subsection{Inflow gas metallicity over cosmic time}
\label{sec:s4:mhaloz}

\begin{figure}
    \centering
    \includegraphics[width=1.0\columnwidth]{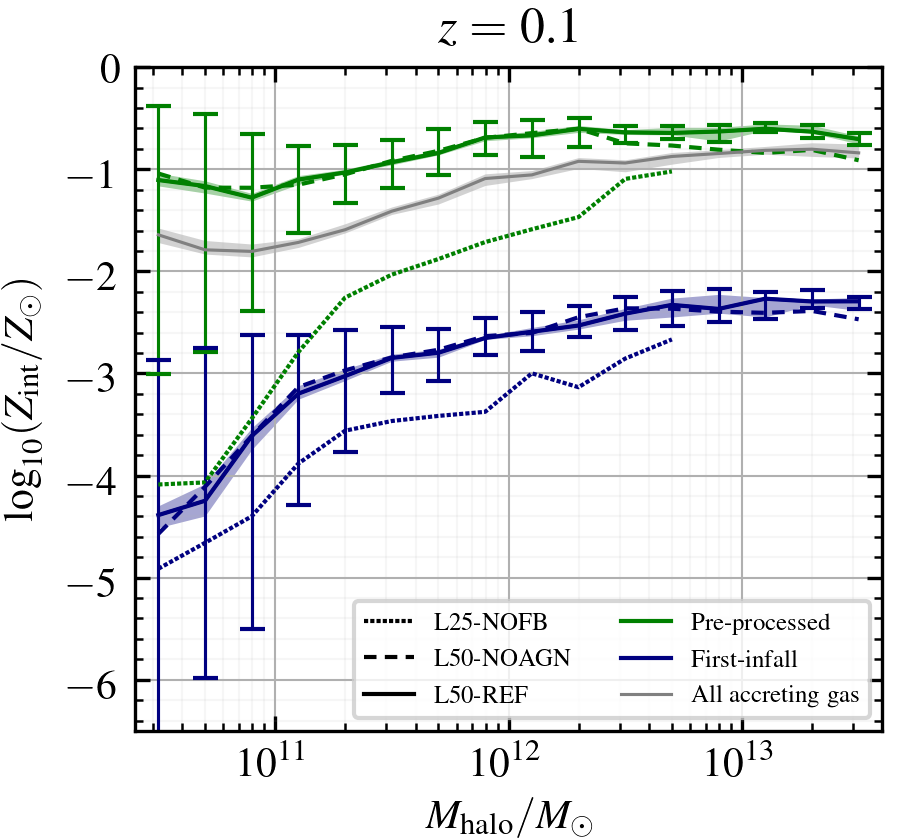}
    \caption{The integrated metallicity (prior to accretion) of halo-accreting gas as a function of halo mass for the history-based inflow modes, excluding the heavily enriched merger-mode. To guide the reader's eye, we also include the integrated metallicity of all accreting gas (essentially a weighted average of the pre-processed and first-infall mode) in the L50-REF run in grey. In this figure we quote the metallicities at $z\approx0$, and consider redshift evolution in Figure \ref{fig:s4:zZ}. The metallicities are shown for multiple \eagle\ runs, namely the L50-REF run (solid lines), the L50-NOAGN run (dashed lines), and the L25-NOFB run (dotted lines). We also include the $16^{\rm th}-84^{\rm th}$ percentile spread (between haloes) in accreting metallicity for L50-REF as error bars, and the bootstrap-generated $95\%$ confidence interval error on the median for L50-REF as shaded regions. }
    \label{fig:s4:massZ}
\end{figure}

\begin{figure}
    \centering
    \includegraphics[width=1.0\columnwidth]{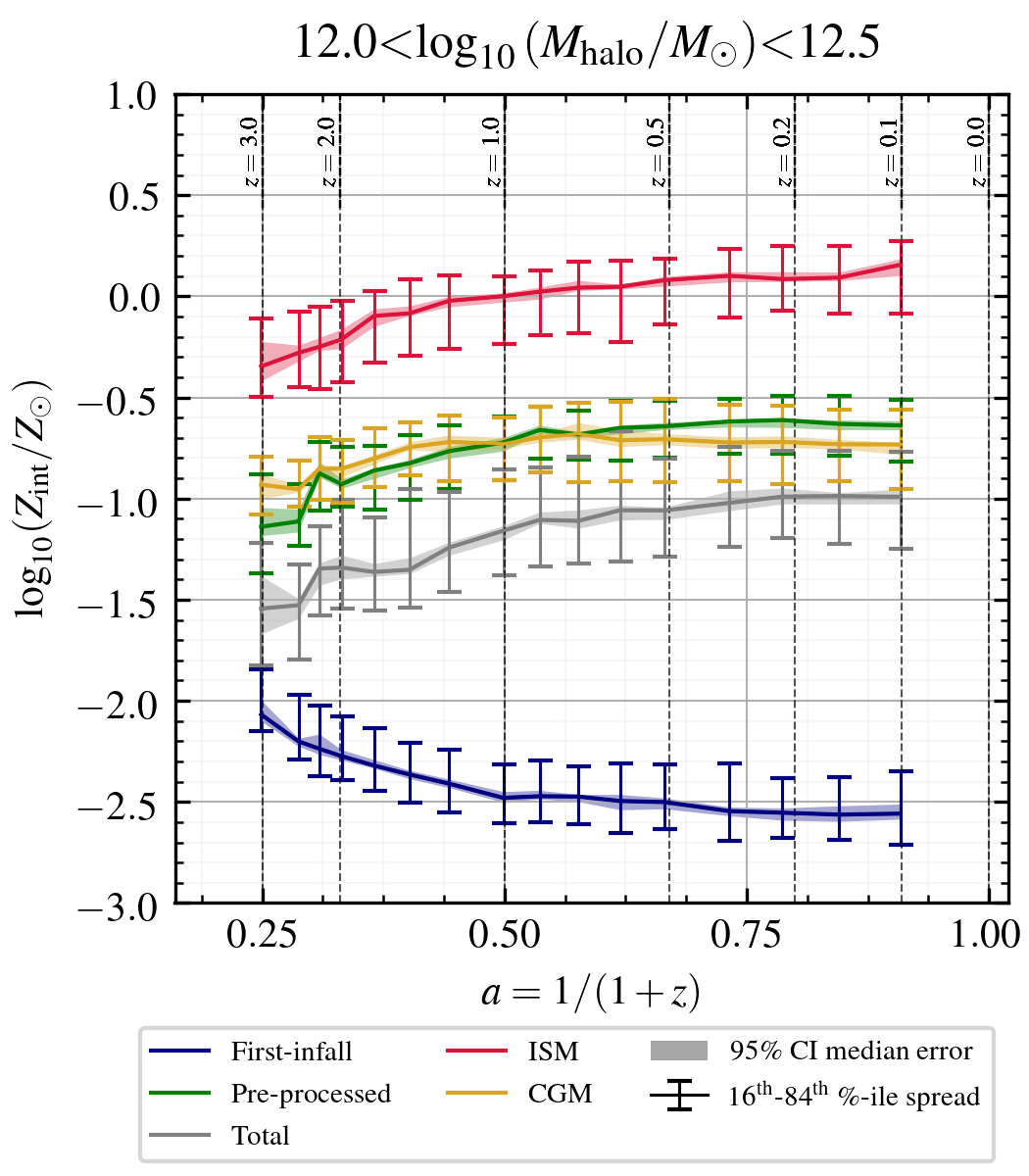}
    \caption{The pre-accretion integrated metallicity of first-infall (blue), pre-processed (green) and all accreting gas (grey) as a function of scale factor for haloes between $10^{12}M_{\odot}$ and $10^{12.5} M_{\odot}$ in L50-REF only. For comparison, in the same mass sample we also include the integrated metal content of halo CGMs (yellow) and central galaxy ISMs (hot pink) as defined in the text. For each line, we include the $16^{\rm th}-84^{\rm th}$ percentile range spread in accreting metallicity as error bars, and the bootstrap-generated $95\%$ confidence interval error on the median as shaded regions. }
    \label{fig:s4:zZ}
\end{figure}

Figure \ref{fig:s4:massZ} shows the median integrated metallicity  (over haloes) of accretion channels as a function of halo mass for first-infall and pre-processed particles at $z\approx0$ (we defer discussion of the redshift evolution of accretion metallicities to Figure \ref{fig:s4:zZ}). The most evident feature is the clear and near-constant disparity between the metallicity of the first-infall and pre-processed mode. From $M_{\rm halo}\approx10^{11}{\rm M}_{\odot}$ to $10^{13.5}{\rm M}_{\odot}$, the metallicity of the first-infall mode is significantly depleted relative to the pre-processed mode by $\approx2-3$~dex, with the pre-processed mode averaging integrated metallicities in the range $-1.5<\log_{10}(Z/Z_{\odot})<-0.5$. The disparity between the first-infall and pre-processed channels is physically intuitive: particles which have not been processed in a halo prior to accretion are far less likely to be enriched compared to those that have previously been accreted onto a halo. We remark to the reader that these history-based classifications of inflow channel yield a much stronger separation in the metallicities of their populations compared to the negligible difference noted between temperature-based (hot- and cold-mode) classifications, which we consequently have not included in Figure \ref{fig:s4:massZ}.

The low metallicities ($\log_{10}(Z/Z_{\odot})\lesssim-2$) of the first-infall channel, together with the low contamination by stellar-feedback affected gas illustrated in Figure \ref{fig:apdx:tmax:channels} give us confidence in our classification scheme. Despite being very metal poor, first-infall particles are not necessarily pristine (i.e., they do not have zero metallicity). Upon investigation, we found that in most haloes, the majority of the first-infall particles {\it do} possess zero metallicity, but the addition of a small sample of slightly enriched particles can drastically increase the value of $Z_{\rm int}$: the sum of all accreted mass in metals divided by the total mass accreted. The finite cadence of the simulation outputs used, as well as the lower mass resolution limit of \velociraptor, may also lead to a small number of particles wrongly being classed as unprocessed. 

We also note a trend of increasing inflow enrichment with increasing halo mass for each of the included accretion channels. Upon investigation, we argue that this is the result of pre-accretion enrichment - where particles are enriched prior to formally entering the halo environment. We find that the metallicity profiles of higher mass haloes are more extended (even relative to their size) compared to less massive haloes, meaning that gas can be enriched before crossing the FOF boundary of larger haloes.

We now consider the influence of sub-grid physics on the enrichment of accreting gas, based on the alternative physics curves shown in Figure \ref{fig:s4:massZ}. Comparing the L50-REF and L50-NOAGN runs, we see fairly similar behaviour over halo mass and redshift, indicating that AGN feedback does not heavily influence accreting metallicities. The exception to this statement is that pre-processed gas in the L50-NOAGN run appears marginally less enriched compared to the L50-REF run for haloes with mass between $10^{12.5}M_{\odot}$ and  $10^{13}M_{\odot}$ at $z\approx0$. The pre-processed accreting gas in this mass regime is mostly recycling gas from the main progenitor (see \citealt{Wright2020}). In L50-REF, some of this gas will likely correspond to the metal-enriched central ISM gas expelled due to AGN feedback, given this mass range is that previously associated with AGN-driven outflows \citep{Davies2019,Oppenheimer2020}. The additional source of metal-rich recycling gas in L50-REF compared to L50-NOAGN potentially offers an explanation for the marginal increase in pre-processed accreting metal content. 

Unlike the L50-NOAGN run, we see very notable changes in accreting metallicities when investigating the L25-NOFB run compared to L50-REF. First-infall and pre-processed accreting gas in the L25-NOFB run is metal-poor relative to accreting gas in the L50-REF and L50-NOAGN runs, the difference being the lack of stellar feedback. This depletion is particularly clear in the pre-processed mode, which shows reduced metal content by up to $\approx 1-2$~dex across the full halo mass range at $z\approx0$ (the metal depletion being most significant for lower mass haloes, $M_{\rm halo}\lesssim10^{12}M_{\odot}$). 

Similar to the influence of enriched AGN-driven outflows we discuss above, we argue that it is the lack of circum-halo enriched matter that drives the depletion of accreting matter in this run. In runs with stellar feedback, out-flowing gas in stellar feedback driven winds normally acts to deliver enriched material to the CGM. In the absence of these outflows, the CGM and gas at the halo interface is seldom enriched by the central galaxy - meaning that both first-infall and pre-processed accreting gas is less likely to have ever been enriched. In other words, we argue that the spatial scale of metal enrichment from galaxies in the L25-NOFB run is more concentrated to the galaxy-scale, and does not influence the larger scales from which halo-accreting gas originates. Comparatively, gas accreting at the halo-scale in L50-REF and L50-NOAGN is much more likely to have previously been enriched.

In Figure \ref{fig:s4:zZ}, we concentrate on the redshift evolution of accreting metallicities in the reference L50-REF run for a MW-like halo mass band between $10^{12}$ and $10^{12.5} M_{\odot}$. We include the first-infall (blue) and pre-processed (green) modes, together with the integrated metallicity of all accreting gas in grey (including the first-infall, pre-processed and merger components). For reference, we also compare these inflow metallicities to halo-bound gas reservoirs, namely the metal content of the CGM (yellow) and central galaxy ISM (hot pink). For the purposes of this work, we follow the definition of \citet{Correa2018} in defining the ISM as gas  within a radius of $0.15\times R_{200}$ of a halo's most bound particle (which we assume to reside in the central galaxy), that have either (a) a non-zero SFR, or (b) are part of the atomic phase of the ISM with $n_{\rm H}>0.1\ {\rm cm}^{-3}$ and $T<10^{5}$~K. To define the CGM, we use all FOF particles outside the central $0.15\times R_{200}$ of the halo, subtracting the particles from any nested subhaloes in the host (so as to avoid including the ISM gas associated with satellite galaxies). 

Previously, galaxy gas metallicities at $M_{\star}\approx10^{10.5}M_{\odot}$ (corresponding roughly to the illustrated halo mass range) have been measured to evolve from $12+\log({\rm O/H})\approx8.3$ at $z=2$ to $\approx8.8$ at $z=0$ in both observations and theory (see e.g. \citealt{Maiolino2008} and \citealt{Collacchioni2018} respectively), implying a notable evolution of $\approx0.5$~dex. Encouragingly, our measurements of ISM metallicities, shown in hot pink, demonstrate a trend consistent with these previous studies - evolving from $\log(Z_{\rm ISM}/Z_{\odot})\approx-0.2$ at $z\approx2$ to $\log(Z_{\rm ISM}/Z_{\odot})\approx0.2$ at $z\approx0.1$ (a $\approx0.4$~dex increase). 

We find CGM metallicities also tend to increase in L50-REF towards late times, with a slightly shallower slope than seen with ISM metallicities - evolving from $\log(Z_{\rm CGM}/Z_{\odot})\approx-0.8{\rm\ to}-0.9$ at $z\approx2$ to $\log(Z_{\rm ISM}/Z_{\odot})\approx0.7$ at $z\approx0.1$ (which is $\approx0.7-0.9$~dex depleted compared to ISM metal content, slightly smaller than the gap reported in observations by \citet{Kacprzak2019} at $\approx1$~dex). The metallicity of pre-processed accreting gas matches average CGM metallicities closely over redshift, with a very slighly steeper gradient. This concordance is consistent with a picture where the pre-processed inflow material had previously been accreted into the CGM of progenitor haloes, and subsequently recycled.

\begin{figure*}
    \centering
    \includegraphics[width=1\textwidth]{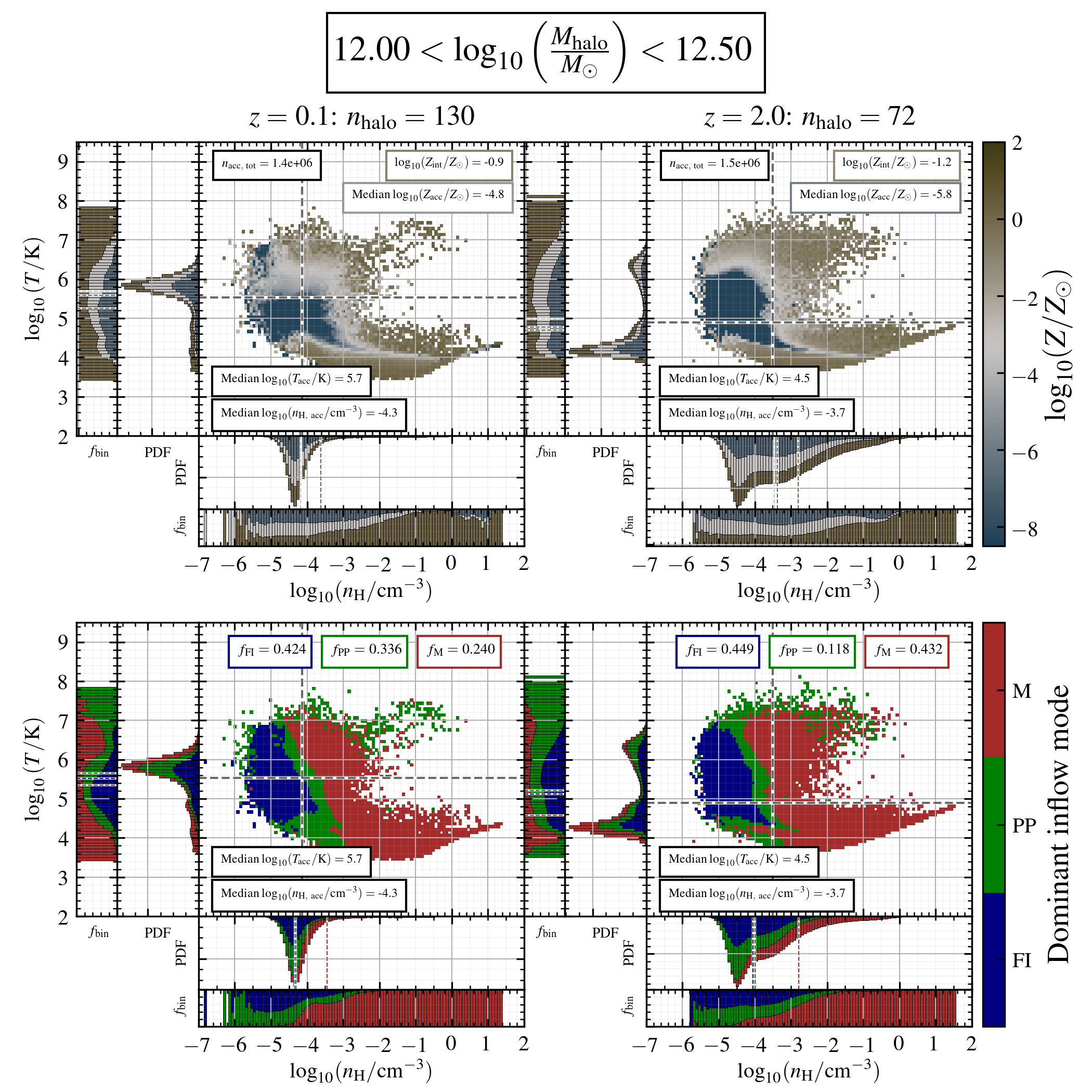}
    \caption{The pre-accretion phase ($n_{\rm H}-T$) plane for gas accreting to haloes of mass $10^{12}<M_{\rm halo}/{\rm M}_{\odot}<10^{12.5}$ at $z\approx0$ (left panels) and $z\approx2$ (right panels). The top panels show the plane coloured by the median accreting metallicity, while the bottom panels show the plane coloured by the dominant inflow mode at a given value of $n_{\rm H}$ and $T$ (``M'' corresponding to merger-origin particles, ``PP'' corresponding to pre-processed particles, and ``FI'' referring to first-infall particles). Also shown are the PDFs of $n_{\rm H}$ and $T$ for accreting particles, in the top panels split into 3 equally log-spaced bins of metallicity and in the bottom panels split into different accretion modes. We also show the fraction of each $n_{\rm H}$ and $T$ bin occupied by each metallicity/inflow mode categorisation. Each panel has an inset at the bottom left indicating the mass-weighted median accreting temperatures and densities, and at the top left of each panel, we quote the total number of gas particles accreted onto haloes in the given mass range. In the top panels, integrated and median accreting metallicities are quoted inset at the top right, while in the bottom panels, the fraction of mass delivered by each inflow mode are quoted at the top of each panel.}
    \label{fig:s4:phasediagram}
\end{figure*}

As seen in Figure \ref{fig:s4:massZ}, the first-infall mode is significantly depleted in metal content compared to the pre-processed mode. The integrated metallicity of this already depleted mode appears to decrease slightly towards $z=0$, from $\log(Z_{\rm int}/Z_{\odot})\approx-2.3$ at $z\approx2$ to $\log(Z_{\rm int}/Z_{\odot})\approx-2.6$ at $z\approx0.1$. This systematically drags down the average integrated metallicity of all accreting gas (grey line) from that of the pre-processed mode by $\approx0.4-0.5$~dex - indicating that, on average, gas accreting to these haloes is metal-depleted by $\approx0.4-0.5$~dex compared to their circum-galactic media, and by $\approx1-1.5$~dex compared to the ISM of their central galaxies. 

Our work with \eagle\ indicates that while very low-metallicity gas ($\log_{10}Z/Z_{\odot}\lesssim10^{-1.5}$) in the CGM is likely to be accreting and on first-infall, gas which has been ejected from a galaxy will likely possess metallicity degenerate with that of returning pre-processed gas. This highlights the importance of kinematic information in observations to delineate between accreting and outgoing material in the CGM of haloes. 

\subsection{Inflow metallicity and the density-temperature plane}
\label{sec:s4:phasediagram}

Figure \ref{fig:s4:phasediagram} illustrates the pre-accretion phase diagram (that is, the density and temperature of gas at the snapshot prior to accretion) of all gas particles accreting to haloes in the mass range $10^{12}<M_{\rm halo}/{\rm M}_{\odot}<10^{12.5}$ for $z\approx0$ (left panels) and $z\approx2$ (right panels). The phase diagram is coloured by the log-space median accreting gas metallicity in the top panels, and the dominant inflow mode in the bottom panels (``M'' corresponding to merger-origin particles, ``PP'' corresponding to pre-processed particles, and ``FI'' referring to first-infall particles).  We note that wherever we calculate averages (means or medians), percentiles, and histograms, we use a mass-weighting procedure which accounts for the mass of each gas particle. 

Focusing on the $z\approx0$ redshift selection (left panels) in Figure \ref{fig:s4:phasediagram}, we note that most accreting particles lie in the density range $10^{-5}{\rm cm}^{-3}<n_{\rm H}<10^{-3}{\rm cm}^{-3}$, and in the temperature range $10^{5}{\rm K}<T_{\rm gas}<10^{7}{\rm K}$. This is a small subset of the full parameter space, and corresponds to the warm/hot phase of the inter-galactic medium (WHIM). In this region of the parameter space we observe intermediate metallicities, $\log_{10}(Z_{\rm acc}/Z_{\odot})\approx-5\ {\rm to}\ -2$, which corresponds to an non-trivial superposition of all $3$ accretion modes (see bottom left panel). 

If we concentrate on the low metallicity population of accreting matter, corresponding to bluer regions in the top panels of Figure Figure \ref{fig:s4:phasediagram}, we see a fairly well defined  distribution in phase space. The majority of low-metallicity accreting gas ($\log_{10}(Z_{\rm acc}/Z_{\odot})\lesssim-4$) occupies the region defined by $10^{-5}<n_{\rm H}/{\rm cm}^{-3}<10^{-3}$ (at the peak of the bulk population) and $10^{4.5}<T_{\rm gas}/{\rm K}<10^{6}$ (slightly cooler than the bulk population). This region of phase diagram is very well correlated with the first-infall accretion mode in the bottom panel, and concurs with our findings in Section \ref{sec:s3:channels} and Figure \ref{fig:s3:fhot} which indicate that first-infall accretion is marginally cooler on average than the pre-processed and merger channels of inflow. 

We also observe a low metallicity "tail" extending towards the equation of state, forming a near linear and tight sequence from $(n_{\rm H},T)=(10^{-4} {\rm cm}^{-3},10^{4.5}{\rm K}$) to $(10^{-1} {\rm cm}^{-3},10^{4}{\rm K})$. This population corresponds to the metal-poor cooling of pristine gas prior to joining the star-forming equation of state. The same low-Z tail in the bottom panel is dominated by the merger-mode of inflow - simply meaning that there are few particles cooling and reaching densities above $n_{\rm H}=10^{-3}{\rm cm}^{-3}$ outside existing haloes (see also \citealt{Schaller2015}, Figure 7).

Extending away from the blue, low-Z population that we discuss above - there is a well defined concentric gradient outwards towards higher metallicities. Solar and super-solar metallicities characerise the accreting gas above $10^{7}$K, and also dominate the population above $n_{\rm H}=10^{-3}{\rm cm}^{-3}$ (with the exception of the aforementioned tight low-Z cooling tail). At both redshifts, we observe a hot and dense population of enriched accreting gas in the region where $n_{\rm H}\gtrsim10^{-2}{\rm cm}^{-3}$ and $T\gtrsim10^{6.5}{\rm K}$. When we investigated the $T_{\rm max}$ values associated with this population, we noted a distinct peak at $T_{\rm max}=10^{7.5}{\rm K}$ - indicating that these particles have been heated by stellar feedback events (this being the prescribed $\Delta T_{\rm SNe}$ in \eagle). In the bottom panels, we find that a combination of pre-processed and merger channels of accretion contribute in this region of the phase diagram. The pre-processed portion likely corresponds to particles which have been ejected from their halo via stellar feedback, and later re-accreted. 

One clear difference between the two redshifts shown is that we find an extension of the low-metallicity population towards higher temperatures at $z\approx2$. The mass-weighted median metallicity of all accreting gas increases from $\log_{10}(Z/Z_{\odot})=-6.1$ at $z\approx2$ to $\log_{10}(Z/Z_{\odot})=-4.2$ at $z\approx0$, where more particles are accreting via the enriched pre-processed mode (see \citealt{Wright2020}). We also find that the overall pre-accretion temperature histogram of accreting gas becomes distinctly bi-modal at $z\approx2$ compared to the mostly uni-modal distribution at $z\approx0$, with the higher redshift snapshot showing a stronger peak at $T_{\rm acc}\approx10^{4.2} {\rm K}$. This lower temperature population correlates with the more dominant cold-mode of gas accretion at higher redshift, and hosts a non-trivial breakdown of metallicities and inflow modes - increasing in metallicity with increasing density, and simultaneously showing a shift from first-infall, to pre-processed, to merger-mode prevalence with increasing density. In addition, the high-$n_{\rm H}$ tail of the density distribution at $z\approx2$ is much more extended compared to late times, where the contribution of merger-mode gas accretion relative to smooth accretion is enhanced \citep{Wright2020} - a natural result from a hierarchical $\Lambda$-CDM mass-growth scenario. 

Our results show that the different accretion modes do preferentially populate different regions of the three dimensional space between density, temperature, and metallicity. However, there is some level of overlap between those regions, and hence by the position of a particle in this three-dimensional space, one can only suggest the most likely accretion mode.

\section{The link between halo-scale gas accretion and observable halo properties}\label{sec:s5}

\begin{figure*}
    \centering
    \includegraphics[width=0.91\textwidth]{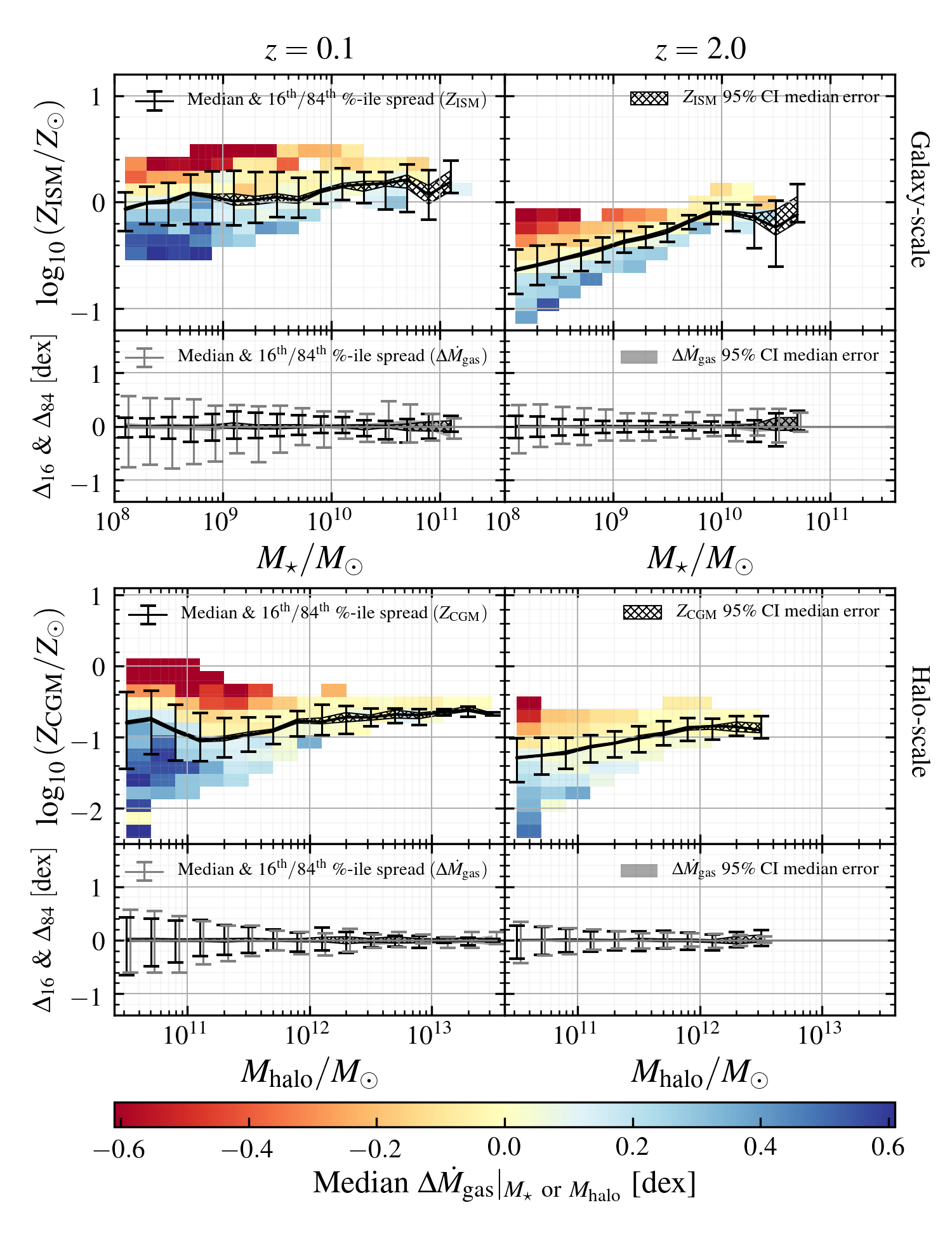}
    \caption{Top 4 panels: The stellar mass - ISM metallicity relation in L50-REF coloured by excess gas accretion efficiency (at fixed stellar mass, $\Delta \dot{M}_{\rm gas}\vert_{M_{\star}}$ - upper panels), and $16^{\rm th}-84^{\rm th}$ percentile range in ISM metallicities and excess gas accretion rates (lower panels). The bins in ISM metallicity are spaced in increments of $0.1$~dex. Lower 4 panels: The halo mass - CGM metallicity relation in L50-REF coloured by excess gas accretion efficiency (at fixed halo mass, $\Delta \dot{M}_{\rm gas}\vert_{M_{\rm halo}}$ - upper panels) and $16^{\rm th}-84^{\rm th}$ percentile range in CGM metallicities and excess gas accretion rates (lower panels).  The bins in CGM metallicity are spaced in increments of $0.2$~dex. Left panels represent the relations at $z\approx0$ and right panels represent the relations at $z\approx2$. The hatched regions show the bootstrap-generated $95\%$ confidence interval  error on the median for L50-REF in each relevant panel, and the parameter space is only coloured where there are at least 5 objects in each 2D bin. }
    \label{fig:s5:massZ}
\end{figure*}
 
\begin{figure*}
    \centering
    \includegraphics[width=0.95\textwidth]{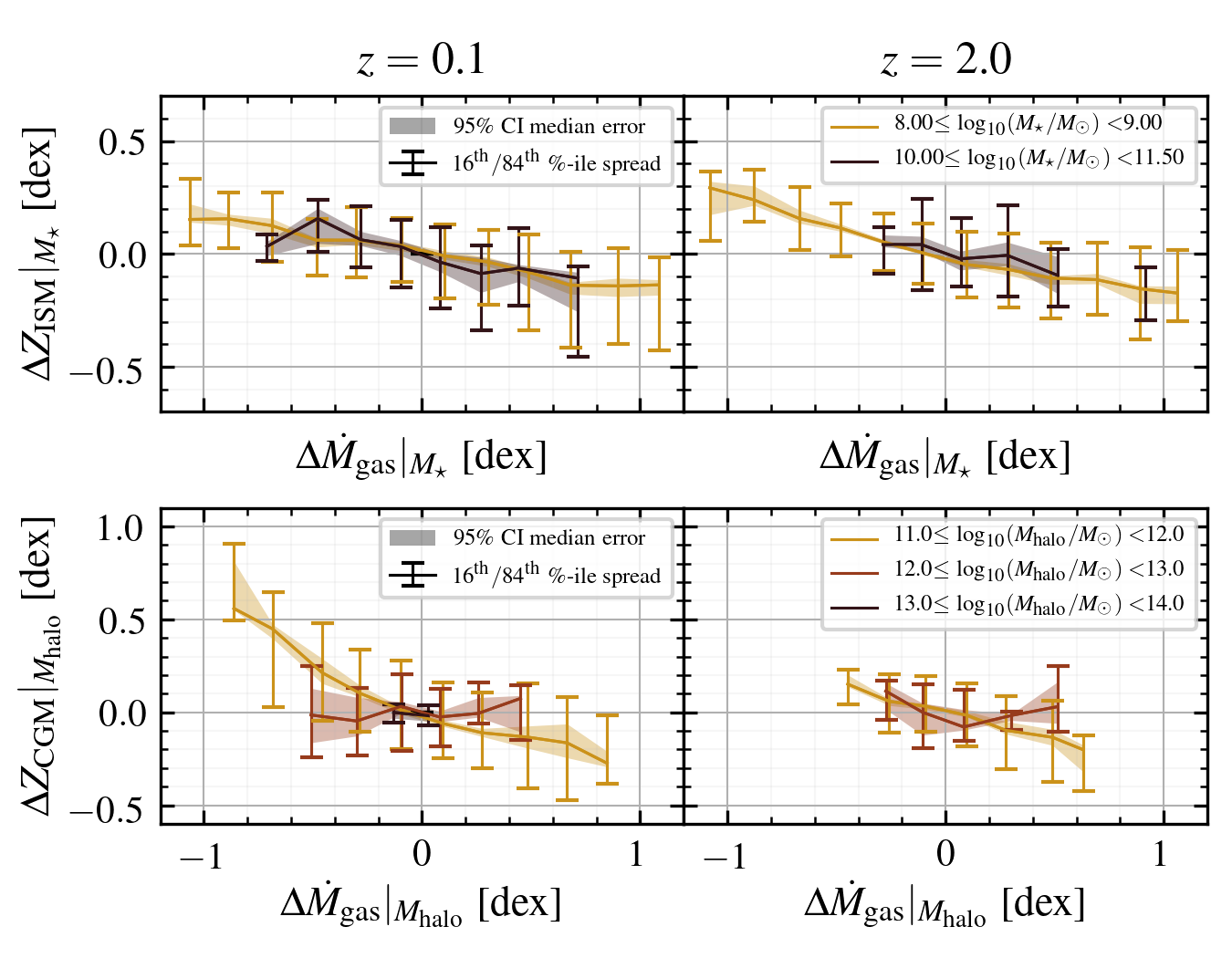}
    \caption{Top panels: the relationship between excess gas accretion rate (at fixed stellar mass, $\Delta \dot{M}_{\rm gas}\vert_{M_{\star}}$) and excess central ISM metallicity ($\Delta Z_{\rm ISM}\vert_{M_{\star}}$) in L50-REF  for 2 bins of central galaxy stellar mass: $\log_{10}(M_{\star}/M_{\odot}) \in [8,9),\ {\rm and\ } [10,11.5)$. Bottom panels: the relationship between excess gas accretion rate (at fixed halo mass, $\Delta \dot{M}_{\rm gas}\vert_{M_{\rm halo}}$) and excess CGM metallicity ($\Delta Z_{\rm CGM}\vert_{M_{\rm halo}}$) in L50-REF for 3 bins of halo mass: $\log_{10}(M_{\rm halo}/M_{\odot}) \in [11,12),\ [12,13),\ {\rm and\ }[13,14)$. Left panels show these relations at $z\approx0$, while right panels show the relations at $z\approx2$. Error bars correspond to $16^{\rm th}-84^{\rm th}$ percentile ranges, and shaded regions show the bootstrap-generated $95\%$ confidence interval  error on the median. }
    \label{fig:s5:deltaZ}
\end{figure*}

 Gas accretion rates have been proposed to be a primary regulator of both galaxy star formation rates (SFRs) and gas metallicities in equilibrium models (e.g. \citealt{Dave2012,Lilly2013}). \citet{DeLucia2020}, using the semi-analytic model GAEA, and  \citet{Collacchioni2019} and \citet{vanLoon2021}, using \eagle, show that gas accretion rates onto galaxies are correlated with the scatter in their predicted stellar mass - ISM metallicity relations (ISM MZR). The physical reasoning behind this modulation is that low-Z, relatively pristine (low-Z) inflow can act to ``dilute'' the metallicity of the ISM. Since gas accretion rates onto galaxies depend strongly on the gas accretion rate onto haloes \citep{Dave2012,Mitchell2020}, we here explore how halo-scale gas accretion rates modulate the properties of their central galaxies and circum-galactic media. In particular, we investigate (i) the ISM MZR and halo mass - CGM metallicity relation (CGM MZR) in \ref{sec:s5:met}, and (ii) the stellar mass - specific SFR (sSFR) relation in \S \ref{sec:s5:ssfr}. 

We frequently use the notation $\Delta x_{\rm i}$ to represent the excess value of quantity $x$ for object $i$ relative to the median. We nominally compute $\Delta x_{i}$ relative to the median of $x$ for a given value of second quantity, $y$ as:
\begin{equation}
\Delta x_{\rm i}\vert_{y}=\log_{10}\left(\frac{x_{i}}{\Tilde{x}\vert_{y_{i}}}\right), 
\label{eq:excess}
\end{equation}
\noindent{where $\Tilde{x}\vert_{y_{i}}$ is the median value of $x$ at $y\approx y_{i}$. Put simply, $\Delta x_{\rm i}\vert_{y}$ describes the log-space excess value of $x$ for object $i$ compared to what would be expected of $x$ from the object's $y$ value. }

\subsection{Mass-metallicity relations}
\label{sec:s5:met}

Here, we investigate the dependence of the (i) ISM MZR and (ii) CGM MZR on halo-scale gas accretion rates using reference \eagle\ physics. We remind the reader that we use \velociraptor\ (and not {\sc SUBFIND}) to find haloes in \eagle, and thus our results are not identical to what would be obtained from the public catalogues outlined in \citet{McAlpine2016}. We find, regardless, that our population statistics agree very well with these data (see Figure 2 of \citealt{Wright2020}, comparing our accretion catalogue to \citealt{Correa2018}). To compute ISM and CGM properties, we adopt the definitions outlined in S\ \ref{sec:s4:mhaloz}.

The top panels of Figure \ref{fig:s5:massZ} show ISM metallicity as a function of $30$~kpc aperture stellar mass (the {\it ISM MZR}) in the L50-REF run for $z\approx0$ and $z\approx2$, with the parameter space coloured by the median gas accretion rate excess for a given stellar mass ($\Delta \dot{M}_{\rm gas}\vert_{M_{\star}}$). These gas accretion rates include the contribution from all inflow modes. The panels below these illustrate the deviation of the $16^{\rm th}$ and $84^{\rm th}$ percentile values for ISM metallicity (black) and the same spread in associated halo-scale gas accretion rates (grey). We note before continuing that the ISM MZR produced by reference \eagle\ physics is is known to be too flat relative to observations below $M_{\star}\approx10^{9.5}M_{\odot}$ \citep{Schaye2015}, however our work focuses on analysing the origin of the scatter in the relation (and not its absolute normalisation). We also note that the scatter in the gas-phase metallicity of high stellar-mass galaxies in \eagle\ is higher than observed due to sampling issues, with such objects typically quenched and therefore containing few star-forming gas particles \citep{Schaye2015}. 

At $z\approx0$ we obtain a fairly flat ISM MZR, increasing slightly from $\log_{10}(Z_{\rm ISM}/Z_{\odot})\approx0$ at $M_{\star}\approx10^{8}M_{\odot}$ to $\approx0.2$ at $M_{\star}\approx10^{11}M_{\odot}$. We find that the scatter in the relation varies slightly with stellar mass, with a peak at $M_{\star}\approx10^{9}M_{\odot}$ (of $\approx0.5$~dex), and minimum at  $M_{\star}\approx10^{10}M_{\odot}$ (of $\approx0.3$~dex). At $z\approx2$, the MZR is steeper, with metallicity increasing from $\log_{10}(Z_{\rm ISM}/Z_{\odot})\approx-0.6$ at $M_{\star}\approx10^{8}M_{\odot}$ to $\approx-0.2$ at $M_{\star}\approx10^{10}M_{\odot}$, flattening above this mass. The spread decreases with stellar mass until just below $M_{\star}\approx10^{10}M_{\odot}$, increasing above this mass range for the most massive galaxies. 

At both redshifts, we find a clear trend with the scatter in ISM metallicity being negatively correlated with halo-scale gas accretion.  This confirms that the ISM MZR modulation by the galaxy-scale gas accretion is in large part driven by the halo scale gas accretion. While we observe this correlation with total accretion rates, we find that the trend remains when using the first-infall or pre-processed modes individually. This is due to the accretion from both these modes being metal depleted relative to the ISM, even in the case of pre-processing (see Figure \ref{fig:s4:massZ}), meaning the gas is able to dilute its metal content. 

Moving to a different scale, the bottom  coloured panels of Figure \ref{fig:s5:massZ} show instead the {\it CGM MZR} in the L50-REF run, again coloured by $\Delta \dot{M}_{\rm gas}\vert_{M_{\rm halo}}$, including the contribution from all modes of inflow. At both redshifts, the metallicity of the CGM is lower than the ISM by $\approx 0.7-1$~dex - this offset being slightly smaller than the $1$~dex offset reported in observations by \citet{Kacprzak2019} (as discussed in \S\ \ref{sec:s4:mhaloz}). At $z\approx0$, the CGM MZR shows a slight minimum at $M_{\rm halo}\approx10^{11}M_{\odot}$ of $Z_{\rm CGM}\approx10^{-1.0}Z_{\odot}$, and above this mass scale, $Z_{\rm CGM}$ increases modestly with halo mass, eventually flattening above $M_{\rm halo}\approx10^{12.5}M_{\odot}$ at $Z_{\rm CGM}\approx10^{-0.8}Z_{\odot}$. At $z\approx2$,  $Z_{\rm CGM}$ increases monotonically with halo mass, from $\approx10^{-1.2}Z_{\odot}$ at a halo mass of $\approx10^{11}M_{\odot}$ to $\approx10^{-0.9}Z_{\odot}$ at $\approx10^{12.5}M_{\odot}$. We remind the reader that that the metallicity of the CGM closely mimics the average pre-processed accreting gas metallicity at both redshifts (see Figure \ref{fig:s4:zZ}), albeit with less variation in $Z_{\rm CGM}$ compared to $Z_{\rm PP}$ at a given halo mass. 

Unlike the ISM MZR, at $z\approx0$ the spread  in $Z_{\rm CGM}$ decreases monotonically with  increasing halo mass from $\approx1$~dex at $M_{\rm halo}\approx10^{11}M_{\odot}$ to  $\approx0.1$~dex at $M_{\rm halo}\approx10^{13}M_{\odot}$. At $z\approx2$ the spread is lower (on average, by $\approx 0.6$~dex) and decreases slightly with increasing halo mass. Focusing on $z\approx0$, for haloes below $10^{12}M_{\odot}$, CGM metal content is strongly correlated with halo-scale gas accretion rates at fixed halo mass - with higher inflow rates associated with lower CGM metallicities. The same trend is evident at $z\approx2$, across an overall smaller spread in $Z_{\rm CGM}$ for a given halo mass. 

We find a very strong gradient in excess gas accretion rate ($\Delta \dot{M}_{\rm gas}\vert_{M_{\rm halo}}$) across the scatter in $Z_{\rm CGM}$ at fixed halo mass, below $M_{\rm halo}\approx10^{12}M_{\odot}$ at $z\approx0$, and below $M_{\rm halo}\approx10^{11.5}M_{\odot}$ at $z\approx2$. In these mass regimes, it is the first-infall mode that dominates the total accretion rate (see Figure \ref{fig:s3:channelefficiencies}) - bringing plentiful low-Z gas (see Figure \ref{fig:s4:massZ}, $Z_{\rm FI}\lesssim10^{-3}Z_{\odot}$) into the CGM, and allowing for its dilution in a similar fashion as discussed in the gas of the ISM.  We explore the mass dependence of CGM dilution further in the discussion around Figure \ref{fig:s5:deltaZ}. 

We explore the driving nature of gas accretion in moderating ISM and CGM metallities more explicitly in Figure~\ref{fig:s5:deltaZ}. In each panel, the x-axes correspond to the excess gas accretion rate (at fixed stellar mass for the ISM, and fixed halo mass for the CGM), and the y-axes in the top (bottom) panels correspond to the excess metallicity of the ISM (CGM) for a given stellar (halo) mass, showing the relationship for a number of bins in stellar (halo) mass. We choose to take two bins in stellar mass, in the ranges $\log_{10}(M_{\star}/M_{\odot})\in[8,9)$ and $[10,11.5)$ respectively, which we refer to ``low-mass'' centrals and ``high-mass'' centrals respectively. These bins are chosen to be consistent with the stellar-mass binning in Figure \ref{fig:s5:deltassfr}, motivated in \S\ \ref{sec:s5:ssfr}. Our bins in halo mass to correspond to the regimes explored in Figure \ref{fig:s3:gasvisualisation} - (i) dwarf-mass haloes, (ii) MW-like haloes, and (iii) group-mass haloes. 

Focusing on the ISM-scale, we find a clear and significant negative correlation between excess total gas accretion and excess ISM metal content for both low- and high-mass central galaxies. The gradient of the relation remains very similar between mass bins, with high-mass galaxies spanning a smaller dynamic range in gas accretion excess, particularly at $z\approx2$. On average, for a factor of 10 increase in gas accretion, we find an associated change in instantaneous ISM metallicity content of $-0.4$~dex (a factor of $\approx2.5$ decrease). 

At the CGM-scale, we find at both redshifts that the only significant trend between gas accretion and CGM metal content exists in the lowest mass bin, $10^{11}M_{\odot}\leq M_{\rm halo}<10^{12}M_{\odot}$, as can be qualitatively seen in Figure~\ref{fig:s5:deltaZ}.  While the correlation only appears significant for the dwarf halo mass range, this halo mass bin exhibits a steeper gradient than we found at the ISM-scale at $z\approx0$. The dynamic range spanned in gas accretion is greatest for this low-mass bin, particularly at $z\approx0$,  aiding the establishment of the anti-correlation above. For a factor of 10 increase in gas accretion efficiency, on average we find a change in CGM metal content of $\approx-0.6$~dex, or a factor of $\approx4$ decrease.

At $z\approx0$ and $z\approx2$, below halo masses of $10^{12}M_{\odot}$, it is the first-infall mode that dominates accretion rates (see Figure \ref{fig:s3:channelefficiencies}) - bringing plentiful low-Z gas into the CGM (see Figure \ref{fig:s4:massZ}), and allowing for its dilution. Upon investigation, we also note that at low halo mass, newly accreted matter corresponds to a significant portion of the CGM mass (e.g. at $z\approx0$ for $10^{11}M_{\odot}$ haloes, $\approx50\%$ of the CGM is newly accreted gas), while in higher mass systems, newly accreted gas tends to make a smaller contribution to the CGM (e.g. at $z\approx0$ for $10^{13}M_{\odot}$ haloes, $\approx25\%$ of the CGM is newly accreted gas). This highlights the dynamic nature of the CGM reservoir, with large fractions of its gas being renewed within a dynamical timescale. 

We remark that in the lower mass range, we also find that the accreting gas is, on average, more metal rich when accretion rates are lower (not shown). However, with the first-infall mode still dominating, these increased metallicities are still below $10^{-3}Z_{\odot}$. Thus, we attribute the increase in $Z_{\rm CGM}$ in low accretion rate haloes to the {\it lack of} pristine inflow, rather than the slight increase in average accretion metal content associated with low accretion rates. 

Above $M_{\rm halo}\approx10^{12}M_{\odot}$, the trend for CGM dilution with increasing accretion rate is less clear at both redshifts. At $z\approx0$ and halo masses above $10^{12}M_{\odot}$, pre-processed accretion rates, on average, exceed first-infall accretion rates (see Figure \ref{fig:s3:channelefficiencies}). The pre-processed mode averages pre-accretion metallicities of $\approx10^{-0.8}Z_{\odot}$ in this mass range (see Figure \ref{fig:s4:massZ}), which corresponds very closely to the measured metallicity of the CGM in this mass range. This means that in this mass range, the effect of CGM dilution by gas accretion may not necessarily be expected - and is indeed not observed.

At $z\approx2$, above $\approx10^{12}M_{\odot}$, we find that accreting metallicities are typically below the metallicity of the CGM (with first-infall still dominating, see Figure \ref{fig:s3:channelefficiencies}), suggesting that we may expect to see a net CGM dilution effect at higher masses for this redshift.  At this redshift, however, we find the spread in CGM metallicities for haloes with mass $\gtrsim 10^{12}M_{\odot}$ is very low, and in the same mass range that the spread in $\Delta \dot{M}_{\rm gas}\vert_{M_{\rm halo}}$ is also extremely small ($\lesssim0.4$~dex, see \citealt{Wright2020} and grey error bars in the bottom right panel of Figure \ref{fig:s5:massZ}). This lack of dynamic range makes it very difficult to establish a trend between the two variables with limited objects at this mass.  

We remark that in the dwarf-mass bin at $z\approx0$, the gradient in excess CGM metallicity is steepest in the regime where excess gas accretion is below the median; i.e. the lack of gas accretion to the halo appears to drive CGM metallicities up, and a surplus of gas accretion decreases CGM metallicities comparatively modestly. In this regime,  haloes with the lowest accretion rates are those being most significantly influenced by stellar feedback (compare L25-NOFB and L50-NOAGN in Figure \ref{fig:s3:channelefficiencies} and \citealt{Wright2020}). In such haloes, there is likely a dual effect occurring: there being less CGM metal dilution due to lack of inflow, but also enhanced metal-rich outflows from feedback - both acting simultaneously to increase the CGM's bulk metal fraction. 

 We also remark that we do not see the same trend for low-mass CGM dilution when using the L25-NOFB run. While accretion rates still correlate with the scatter in the ISM MZR in this run, this is not the case for the CGM, where we find that the circum-galactic gas is minimally enriched at low halo mass ($Z_{\rm CGM}<10^{-2}Z_{\odot}$). The lack of CGM metal content is a result of the lack of enriching stellar-feedback driven outflows in this run, with the enriched gas being mostly confined to the central galaxy. Thus, on average, accreting gas is of similar metallicity to the CGM even at dwarf halo masses, and gas inflow cannot act to dilute its bulk metal content. 

 Our results in \S \ref{sec:s5:met} indicate that halo-scale accretion acts to modulate the ISM MZR (by altering galaxy-scale accretion rates) and influences, arguably even more strongly, the metal content of the CGM in low-mass haloes. This highlights the fact that the CGM is a very dynamic reservoir, particularly in low-mass systems.

\subsection{The star-forming main sequence}
\label{sec:s5:ssfr}

\begin{figure*}
    \centering
    \includegraphics[width=0.92\textwidth]{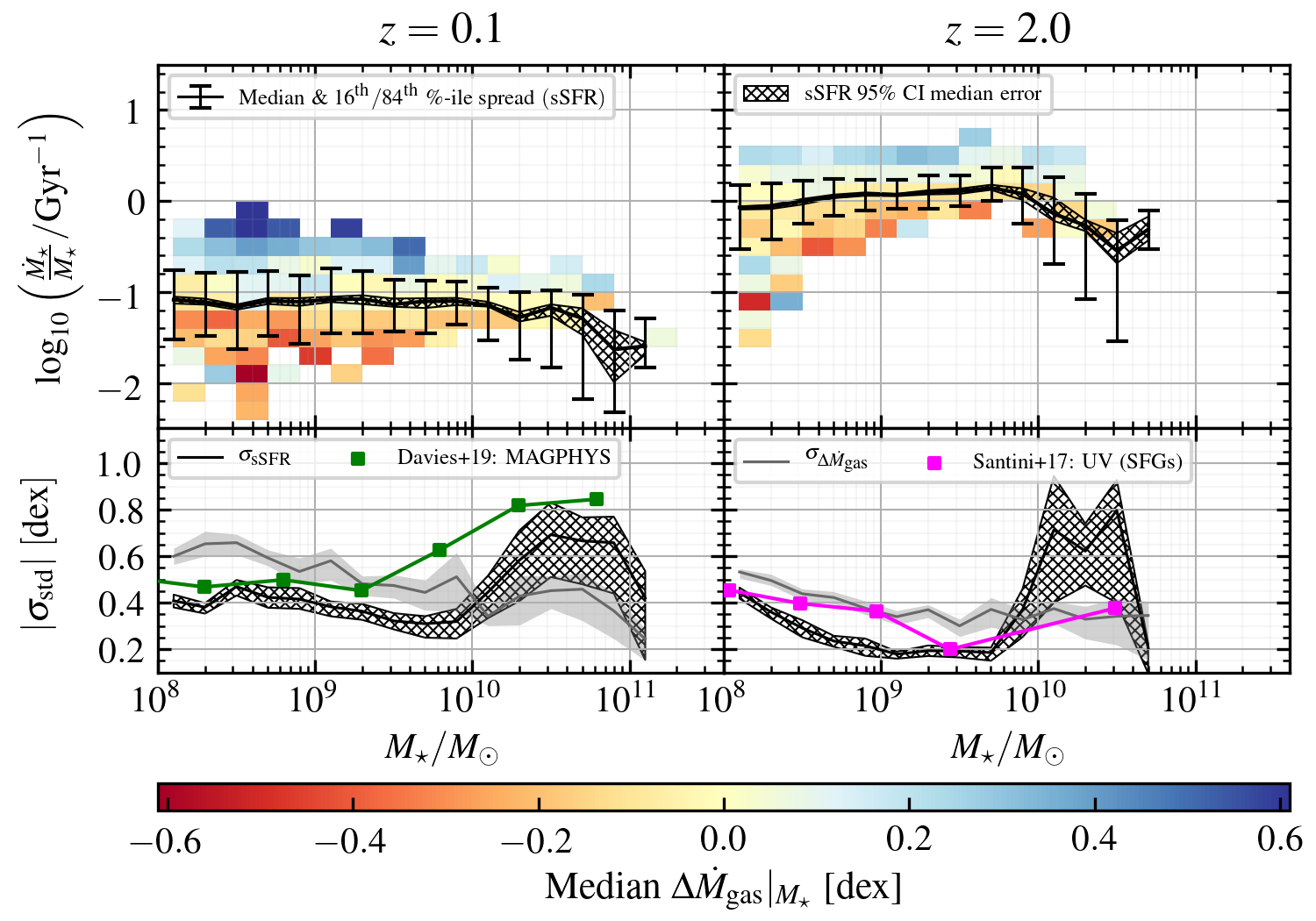}
    \caption{Top panels: The specific star formation rate (${\rm sSFR}$; $\dot{M}_{\star}/M_{\star}$) - stellar mass plane, coloured by excess gas accretion efficiency ($\Delta \dot{M}_{\rm gas}$). The bins in sSFR are spaced in increments of $0.2$~dex. Bottom panels: The standard deviation in sSFR as a function of stellar mass (black) together with observational constraints from \citet{Santini2017} at $z\approx2$ and \citet{DaviesL2019} at $z\approx0$. In the bottom panels, we also include the standard deviation in excess gas accretion rate at fixed stellar mass (grey). Left panels correspond to a selection at $z\approx0$, while right panels correspond to a selection at $z\approx2$. Error bars show the $16^{\rm th}-84^{\rm th}$ percentile range in sSFR. In the top panels, hatched regions show the bootstrap-generated $95\%$ confidence interval error on the sSFR median, while in the bottom panels, the hatched regions show the bootstrap-generated $95\%$ confidence interval error on the sSFR standard deviation ($\sigma$). Similarly, the grey shaded regions in the bottom panels show the bootstrap-generated $95\%$ confidence interval error on $\sigma_{\Delta \dot{M}_{\rm gas}}$ as a function of stellar mass.}
    
    \label{fig:s5:massssfr}
\end{figure*}

In this section, we investigate the dependence of central galaxy SFRs on halo-scale gas accretion rates in \eagle. We focus specifically on the ${\rm sSFR}\equiv \dot{M}_{\star}/M_{\star}$ - stellar mass plane (Figure \ref{fig:s5:massssfr}) to explore this dependence. 

\begin{figure*}
    \centering
    \includegraphics[width=0.95\textwidth]{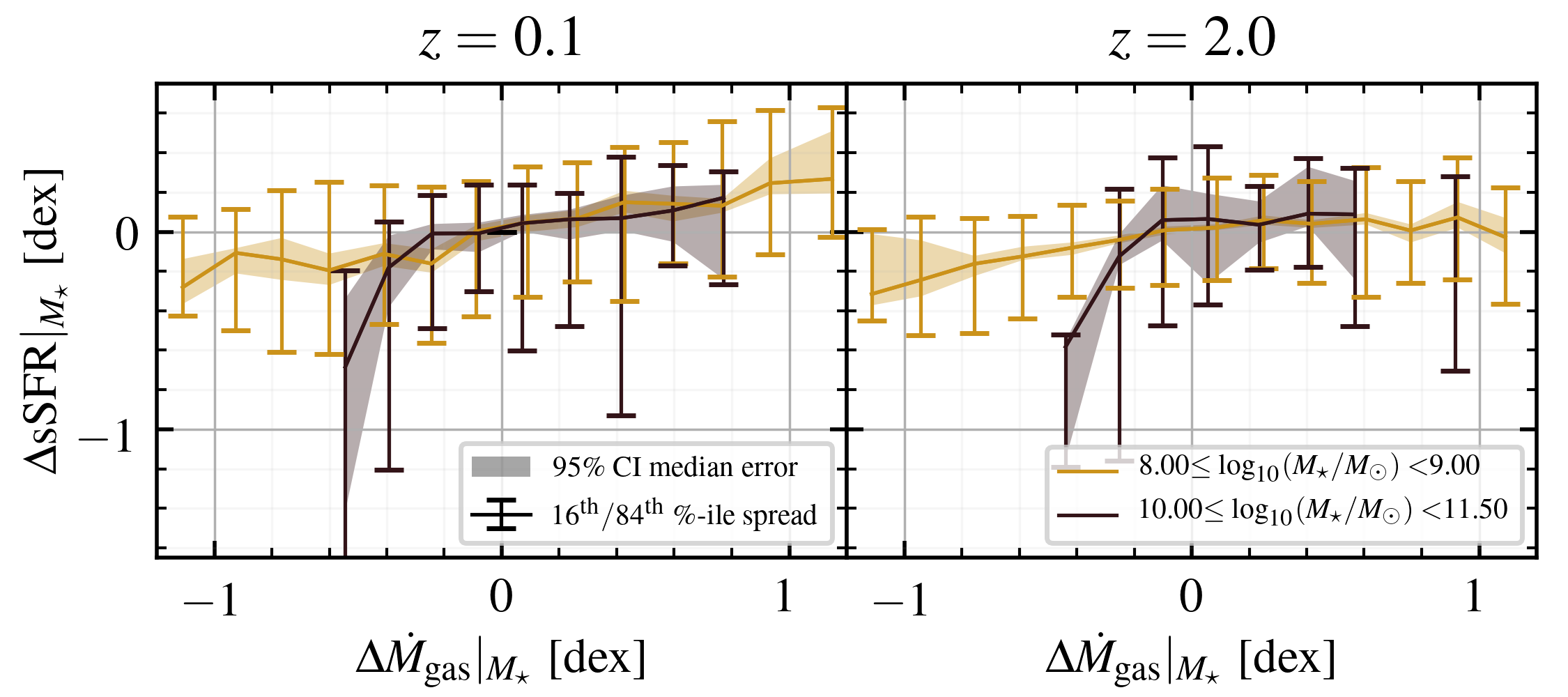}
    \caption{The relationship between excess gas accretion rate ($\Delta \dot{M}_{\rm gas}\vert_{M_{\star}}$) and excess specific star formation rate ($\Delta {\rm sSFR}\vert_{M_{\star}}$) in L50-REF for 2 bins of central galaxy stellar mass: $\log_{10}(M_{\star}/M_{\odot}) \in [8,9),\ {\rm and\ } [10,11.5)$. The left panel shows the relations at $z\approx0$, while right panel shows the relations at $z\approx2$. Error bars correspond to $16^{\rm th}-84^{\rm th}$ percentile ranges, and shaded regions show the bootstrap-generated $95\%$ confidence interval  error on the median. }
    \label{fig:s5:deltassfr}
\end{figure*}

Figure \ref{fig:s5:massssfr} shows the stellar mass - sSFR plane in L50-REF at $z\approx0$ and $z\approx2$, coloured by $\Delta \dot{M}_{\rm gas}\vert_{M_{\star}}$, which includes the contribution from all modes of inflow. To generate our sSFR main sequence (SFMS), we use all central galaxies in haloes with mass $M_{\rm halo}\gtrsim10^{9}M_{\odot}$ and non-zero SFRs. We remark that the normalisation and shape of our SFMS is consistent with that originally presented in \citet{Schaye2015} using the public {\sc subfind} catalogues. At $z\approx0$, sSFRs are relatively flat with stellar mass (at $\approx 10^{-1}\ {\rm Gyr}^{-1}$) up to $M_{\star}\approx10^{10}M_{\odot}$, above which there is a downturn due to the growing passive population.  

At $z\approx0$ and at fixed stellar mass, central galaxies that lie above (below) the SFMS are significantly more likely to see increased (reduced) halo-scale total gas accretion rates, as evidenced by the clear colour gradient across the scatter associated with the SFMS. This trend is still present but less pronounced at $z\approx2$, where the spread in sSFRs and gas accretion rates are smaller.  

The bottom panels of Figure \ref{fig:s5:massssfr} show the standard deviation in the SFMS ($\sigma_{\rm sSFR}$) as a function of stellar mass, together with a selection of observational datasets \citep{Santini2017,DaviesL2019}. Recent observational work has highlighted that $\sigma_{\rm sSFR}$ against $M_{\star}$ has a distinct 'U' shape, with a minimum spread intermediate stellar mass, $M_{\star}\approx10^{9.25}M_{\odot}$ (see Figure~$7$ in \citealt{DaviesL2019}). This has been supported qualitatively with the \eagle\ simulations in \citealt{Katsianis2019}  and the {\sc Shark} semi-analytic model \citep{Lagos2018b} - who find enhanced diversity in the SFRs of low- and high-mass systems (driven by stellar and AGN feedback, respectively), with a minimum $\sigma_{\rm sSFR}$ at intermediate stellar mass. Our results agree with \citet{Katsianis2019}, and demonstrate a minimum in $\sigma_{\rm sSFR}$ at intermediate stellar mass, between $10^{9}M_{\odot}$ and $10^{10}M_{\odot}$, for both $z\approx0$ and $z\approx2$. 

In the bottom panels, we also include the standard deviation in $\Delta \dot{M}_{\rm gas}\vert_{M_{\star}}$ (grey), to draw attention to its behaviour relative to $\sigma_{\rm sSFR}$ over stellar mass. While we expect a different normalisation, we find that $\sigma_{\Delta \dot{M}_{\rm gas}}$ decreases monotonically with stellar mass at both redshifts and does not display the same 'U' shape over stellar mass as $\sigma_{\rm sSFR}$ does. We explore this further in Figure \ref{fig:s5:deltassfr}, which shows galaxy excess sSFR (${\rm \Delta sSFR}\vert_{M_{\star}}$) as a function of excess halo gas accretion rates (at fixed stellar mass, $\Delta \dot{M}_{\rm gas}\vert_{M_{\star}}$) in two bins of stellar mass. We choose these bins to capture the physics behind the flaring of $\sigma_{\rm sSFR}$, which is most significant at low, $\log_{10}(M_{\star}/M_{\odot})\in[8,9)$, and high,  $[10,11.5)$ stellar masses.

Concentrating on $z\approx0$, we see that each mass bin exhibits a positive correlation between ${\rm \Delta sSFR\vert_{M_{\star}}}$ and $\Delta {\rm \dot{M}_{\rm gas}}\vert_{M_{\star}}$, with the dynamic range in $\Delta {\rm \dot{M}_{\rm gas}}\vert_{M_{\star}}$ being greatest for the low-mass sample. In the low-mass central sample, we find that a factor of 10 increase in gas accretion excess on average leads to an increase in excess sSFR of $\approx0.4$~dex, or a factor of $\approx2.5$ (this being true for haloes with both negative and positive gas accretion excess). In the high-mass central sample at $z\approx0$, $10^{10}M_{\odot}\leq M_{\star}< 10^{11.5}M_{\odot}$, we observe a steeper slope at sub-median gas accretion excess values, with a $0.5$~dex increase in gas accretion rates resulting in a $\approx1$~dex increase in central sSFR. The trend in the high mass sample is less obvious above median gas accretion rates, and traces the lower mass sample over a lessened dynamic range in $\Delta {\rm \dot{M}_{\rm gas}}\vert_{M_{\star}}$. 

Interestingly, we remark that the same relation in the L50-NOAGN run (not shown) does not show the same steep gradient at low accretion rates for the high stellar mass bin, suggesting that the spread in sSFR is physically linked to reduced gas accretion onto the galaxy in this mass regime due to AGN activity.  \citet{Wright2020} showed that the inclusion of AGN feedback in \eagle\ leads to an average decrease in gas accretion rates onto haloes of $\approx 30$\% between $M_{\rm halo}\approx10^{12}M_{\odot}$ and $10^{12.5}M_{\odot}$, reducing the capability of galaxies to replenish their ISM and continue star formation.  In massive \eagle\ haloes, AGN feedback acts as a maintenance mode, preventing cooling flows onto galaxies \citep{Bower2017}. This means that even though accretion onto haloes can still take place in the L50-REF run, much of the accreting gas simply will not enter the central galaxy. This is shown explicitly in \citet{Correa2018}, where galaxy-scale accretion rates are reduced by $\approx0.5$~dex (or $\approx300\%$) with the inclusion of AGN feedback at $M_{\rm halo}\approx10^{12.5}M_{\odot}$, compared to the relatively modest reduction in halo-scale accretion rates. This leads to the significant reduction in ${\rm \Delta sSFR}\vert_{M_{\star}}$ in the high stellar mass bin when $\Delta \dot{M}_{\rm gas}\vert_{M_{\star}}<0$. 

At $z\approx2$, the modulating effect of $\Delta \dot{M}_{\rm gas}\vert_{M_{\star}}$ on $\Delta {\rm sSFR}\vert_{M_{\star}}$ is still present, but less clear (as also noted in  Figure \ref{fig:s5:massssfr}). In general, variation about the SFMS is lower at $z\approx2$ compared to $z\approx0$, with a minimum in $\sigma_{\rm sSFR}$ of $\approx0.15$~dex and $\approx0.3$~dex respectively. We also note that the high-mass sample only contains $\approx100$ galaxies at $z\approx2$, providing limited statistical power. We observe a positive trend between $\Delta \dot{M}_{\rm gas}\vert_{M_{\star}}$ and $\Delta {\rm sSFR}\vert_{M_{\star}}$ for the low-mass sample, and a similar steepening for the high-mass sample at sub-median accretion rates. 

Our work provides context for the result presented by \citet{Katsianis2019}, which demonstrated that stellar and AGN feedback are responsible for the increase in $\sigma_{\rm sSFR}$ at low and high stellar masses respectively. While we don't observe a flaring of halo-scale $\sigma_{\dot{M}_{\rm gas}}$ at high stellar masses like $\sigma_{\rm sSFR}$, we note that the effect of AGN is to primarily suppress gas accretion to the central galaxy, rather than the halo itself \citep{Correa2018}. We expect that in their respective mass regimes, stellar and AGN feedback play a dual role in modulating galaxy-scale SFRs by (i) removing gas eligible for star formation (e.g. see \citealt{Mitchell2019,Davies2019}), and (ii) preventing further gas inflow onto haloes \citep{Wright2020}. The galaxies associated with low accretion rates are those being most strongly influenced by feedback, and thus subject to both of these modulating processes. Our halo-scale accretion rate measurements indicate that feedback-induced modulation of gas accretion is one of the physical driving forces that regulate galaxy SFRs.

\section{Discussion and summary}\label{sec:conclusion}
In this paper, we present measurements of the diverse properties of gas accreting to haloes  (namely its spatial distribution, metallicity, density and temperature) in the \eagle\ simulations, based on the methods outlined in \S\ \ref{sec:methods} and \citet{Wright2020}. We decompose the accreting gas into a number of contributing channels based on (a) the history of inflow particles (rows 2-4 of Table \ref{tab:s2:classification}): (i) first-infall mode - inflow particles which have never been identified as part of a halo in the past; (ii) pre-processed mode - particles which had been processed in a halo beforehand, but were most recently accreted from the field; and (iii) merger-mode - particles which were accreted onto a halo which, at the previous snapshot, were part of another halo. In addition to these 3 history-based accretion modes, we also decompose the same (non-merger/smooth) accreting gas into (b) a hot- and cold-mode, based on a post-accretion (post-shock) temperature cut of $10^{5.5}{\ \rm K}$ (rows 5-6 of Table \ref{tab:s2:classification}). 

In \S\ \ref{sec:s3:channels}, we focus on the temperature of gas accreting to haloes, specifically the correspondence between the history-based classification of inflow channels and the hot- and cold- modes of gas accretion. With L50-REF (reference physics), we find that the hot fraction of first-infall gas is lower than the pre-processed mode for haloes in the mass range between $10^{11.5}M_{\odot}$ and $10^{12}M_{\odot}$ at $z\approx0$ and $z\approx2$ - meaning that unprocessed gas is preferentially cold compared to recycled, pre-processed gas.  

In \S\ \ref{sec:s3:spatial}, we compare the spatial characteristics of accreting gas (visualised in Figure \ref{fig:s3:gasvisualisation}), split by the aforementioned inflow channels. We use the ``covering fraction'' of accreting gas ($f_{\rm cov}$, defined in \S\ \ref{sec:s2:spatial}) to quantify how isotropic ($f_{\rm cov}\to1$) or filamentary  ($f_{\rm cov}\to0$) accreting gas is prior to entering a FOF halo. In L50-REF, we find that pre-processed inflow is significantly more isotropic than the first-infall mode, with $z\approx0$ $f_{\rm cov}$ values of approximately $80\%$ and $60\%$ respectively. We also find that hot-mode inflow is similarly more isotropic than cold-mode inflow, with $z\approx0$ $f_{\rm cov}$ values of approximately $80\%$ and $50\%$ respectively -- the temperature-based inflow channels showing a slightly stronger separation in $f_{\rm cov}$ than the history-based channels. The disparity in covering fractions is even greater in the L50-NOAGN run ($\approx40\%$) due to increased hot-mode covering fractions, where we argue that accreting gas can occupy a greater fraction of the virial sphere in the absence of AGN-driven outflows. 

In \S\ \ref{sec:s4:mhaloz}, we explore the metallicity of gas accreting to haloes as a function of halo mass and cosmic time. We find that there is a very clear disparity between the chemical enrichment of accreting first-infall and pre-processed gas across all halo masses, with $z\approx0$ metallicities in L50-REF of $\log_{10}Z_{\rm FI}/Z_{\odot}\approx-2.5$ and $\log_{10}Z_{\rm PP}/Z_{\odot}\approx-0.5$ respectively at $M_{\rm halo}\approx10^{12}M_{\odot}$ (Figure \ref{fig:s4:massZ}). We also find that gas accreting to haloes in the L25-NOFB run is lower in metal content than accreting gas in L50-REF, which appears to be the result of reduced feedback-driven enrichment. In haloes between $10^{12}M_{\odot}$ and $10^{12.5}M_{\odot}$ (MW-like mass), the metal content of pre-processed accreting gas and existing CGM reservoirs are very similar, and grow very closely over cosmic time (within $0.1-0.2$ dex in the range $\log_{10}Z/Z_{\odot}\approx-1.0\ {\rm to}\ -0.5$, see Figure \ref{fig:s4:zZ}). This highlights the degeneracy between accreting and outgoing gas when their enrichment is considered in isolation. ISM metallicities are systematically enhanced compared to CGM metallicities by $\approx0.7-0.9$ dex, with values in the range $\log_{10}Z/Z_{\odot}\approx-1.0\ {\rm to}\ -0.5$, in rough quantitative agreement with the observations of \citet{Kacprzak2019}. Observations of gas accreting to galaxies likely corresponds to this CGM gas rather than first-infall accreting IGM gas, which we predict to be further metal depleted by $\approx2$ dex.

We show the metallicity of infall particles and their likely inflow mode as a function of position on the density-temperature phase plane in Figure \ref{fig:s4:phasediagram}, for haloes in the MW-like mass band. Low-metallicity accreting gas preferentially occupies the low-density regime of the phase diagram ($n_{\rm H}\lesssim10^{-3}{\rm cm}^{-3}$), with a small cooling tail extending towards the \eagle\ imposed equation of state. High-metallicity gas tends to occupy the outskirts of the phase space distribution, with a particularly prominent hot and dense ($T\approx10^7 {\rm K}$, $n_{\rm H}\approx10^{-2}-10^{1}\ {\rm cm}^{-3}$) population at $z\approx2$. We find that this population corresponds primarily to recycling or merging gas that was recently subject to stellar feedback induced heating. 

In general, we find that the history-based classification of accreting gas (specifically, whether the gas has been processed in a halo previously) is a very good predictor of its metallicity; while the classification of accreting gas based on post-shock temperature better predicts the gases' spatial properties.

In \S\ \ref{sec:s5}, we investigate the influence of halo-scale gas accretion on the properties of halo circum-galactic media, and the ISM of their central galaxies. In \S\ \ref{sec:s5:met}, we show that the gas inflow rates to low mass haloes ($M_{\rm halo}\lesssim10^{12}M_{\odot}$) play a role in determining the metallicity of a halo's CGM. This is analogous to the driving force of galaxy-scale inflow, which is shown to regulate the scatter in the galaxy-scale stellar mass - ISM metallicity relation (previously explored in the context of \eagle\ by \citealt{Collacchioni2019} and \citealt{vanLoon2021}). The modulation of CGM metallicity is strongest in this mass regime for several reasons: (i) accreting metallicities are lower, enhancing the metal ``dilution'' effect; (ii) the accreted gas constitutes a larger percentage of the gas reservoir ($\approx50\%$ of CGM gas in $10^{11}M_{\odot}$ haloes has been accreted within the last dynamical timescale, compared to $\approx 25\%$ in $10^{13}M_{\odot}$ haloes); and (iii) there is less scatter in CGM metallicities in higher mass haloes. This highlights the dynamic nature of the CGM as a reservoir, particularly below masses of $10^{12}M_{\odot}$. 

Finally, in \S\ \ref{sec:s5:ssfr}, we explore the influence of halo-scale gas accretion on the scatter of the star-forming main sequence in \eagle. The characteristic 'U' shape in $\sigma_{\rm sSFR}$ over stellar mass, described in observations in \citet{Davies2019} and explored in \eagle\ in \citet{Katsianis2019}, has previously been explained by stellar feedback and AGN feedback in the low and high stellar mass regimes respectively. We show that the central galaxies in haloes experiencing low gas accretion rates (driven by stellar and AGN feedback, see \citealt{Wright2020}) preferentially sit below the star-forming main sequence. Thus, we find that the preventative influence of these feedback mechanisms may offer a natural explanation for variation in central galaxy SFRs, and the consequent flaring of $\sigma_{\rm sSFR}$ at both stellar mass extremes. 

We remind the reader that the sub-grid models included in \eagle\ have been calibrated to match $z=0$ observed galaxy masses and sizes, leaving the remaining parameter space of simulation outputs with predictive value. Our findings show that in \eagle, gas inflow onto haloes leaves a clear imprint on the observable properties of these haloes and their central galaxies, and that the characteristics of the accreting gas are closely tied to its history. These predictions from \eagle\ are physically self-consistent, and align well with observational inferences. Nonetheless, our study represents a field seldom explored directly in observational and theoretical literature, and we hence remark that it is difficult to ascertain the dependence of our quantitative results on the model used. 

Our findings repeatedly highlight the dynamic nature of the CGM: its properties set by the continuous interplay of inter-galactic  gas inflow and galaxy-driven baryonic feedback processes. This makes the CGM of galaxies a particularly pertinent location to study and constrain different aspects of the baryon cycle, sitting at the interface between cosmological and galactic scales. \citet{Mitchell2019} show that there is still considerable uncertainty between models regarding the spatial scale of outflows and recycling within and beyond the CGM. \eagle\ suggests that outflows due to stellar feedback are driven to large radii ($\gtrsim R_{\rm 200}$), in the process entraining a significant amount of CGM gas. Comparatively, the Illustris-TNG and {\sc fire} models suggest a scenario where outflows are not driven as far, with lower baryon ejection rates at the halo-scale than at the galaxy-scale (ultimately leading to higher halo-scale baryon fractions in TNG compared to \eagle, e.g. \citealt{Davies2019}). 

While the models may disagree on the scale of recycling, \citet{Peroux2020} show at $M_{\star}\approx10^{10.5}M_{\odot}$ and an impact parameter of $b\approx100$~kpc (slightly shy of $R_{200}$) that \eagle\ and TNG produce similar mass flow rates as function of azimuthal angle; with inflows and outflows dominating the major and minor axes of galaxies respectively. The same study showed that metal-poor (rich) gas preferentially exists on the major (minor) axis of galaxies at $b\approx100$~kpc in \eagle\ and Illustris-TNG, however given the different recycling scenarios, the agreement between models will likely be a function of impact parameter. As mentioned above, we show that pre-processed gas accretion to haloes in \eagle\ closely mimics the {\it integrated} metallicity of the CGM. Observations, however, have shown that sight-line measurements of CGM metallicity can vary greatly in the same halo (up to $\approx2$~dex, e.g. \citealt{Lehner2013,Prochaska2017,Zahedy2019}). Investigating the scale of metal-enrichment in simulated CGM, particularly studying how angular {\it and} radial variations would manifest in discrete sight-line observations, could constitute a promising test of model accuracy. Upcoming absorption observations of the CGM (using state-of-the-art facilities such as VLT/MUSE and Keck/KCWI) will form an ideal statistical test-bed for such predictions, as well as those that we present in this paper.

\section*{Acknowledgements}
The authors thank the referee for their helpful comments that improved the clarity of this paper. We also thank Dr. Peter Mitchell and Dr. Joop Schaye for their helpful input and suggestions. RW is funded by a Postgraduate Research Scholarship from the University of Western Australia (UWA). CL is funded by the ARC Centre of Excellence for All Sky Astrophysics in 3 Dimensions (ASTRO 3D), through project number CE170100013. CL also thanks the MERAC Foundation for a Postdoctoral Research Award. CP acknowledges the support ASTRO 3D. This work made use of the supercomputer OzSTAR, which is managed through the Centre for Astrophysics and Supercomputing at Swinburne University of Technology. This super-computing facility is supported by Astronomy Australia Limited and the Australian Commonwealth Government through the National Collaborative Research Infrastructure Strategy (NCRIS). The \eagle\ simulations were performed using the DiRAC-2 facility at Durham, managed by the ICC, and the PRACE facility Curie based in France at TGCC, CEA, Bruyeres-le-Chatel. 

The authors used the following software tools for the data analysis and visualisation in the paper: 
\begin{itemize}
    \item {\fontfamily{pcr}\selectfont python3} \citep{vanRossum}
    \item {\fontfamily{pcr}\selectfont numpy} \citep{Harris2020}
    \item {\fontfamily{pcr}\selectfont  matplotlib} \citep{Hunter2007}
    \item {\fontfamily{pcr}\selectfont  yt} \citep{Turk2011}
\end{itemize}

\section*{Data Availability}

Particle data from the set of the \eagle\ runs used for our analysis is publicly available at \url{http://dataweb.cosma.dur.ac.uk:8080/eagle-snapshots/} - specifically the L25-REF, L50-REF, L50-NOAGN, and L25N752-RECAL runs. All \velociraptor-generated halo catalogues and  \treefrog\ merger trees are available upon request from the corresponding author (RW). The code we used to generate accretion rates to haloes is available at \url{https://github.com/RJWright25/CHUMM}.



\bibliographystyle{mnras}
\bibliography{mnras_template.bib} 


\appendix

\section{Maximum temperature of accreting gas particles}\label{sec:apdx:tmax}

\begin{figure*}

    \centering
    \includegraphics[width=0.9\textwidth]{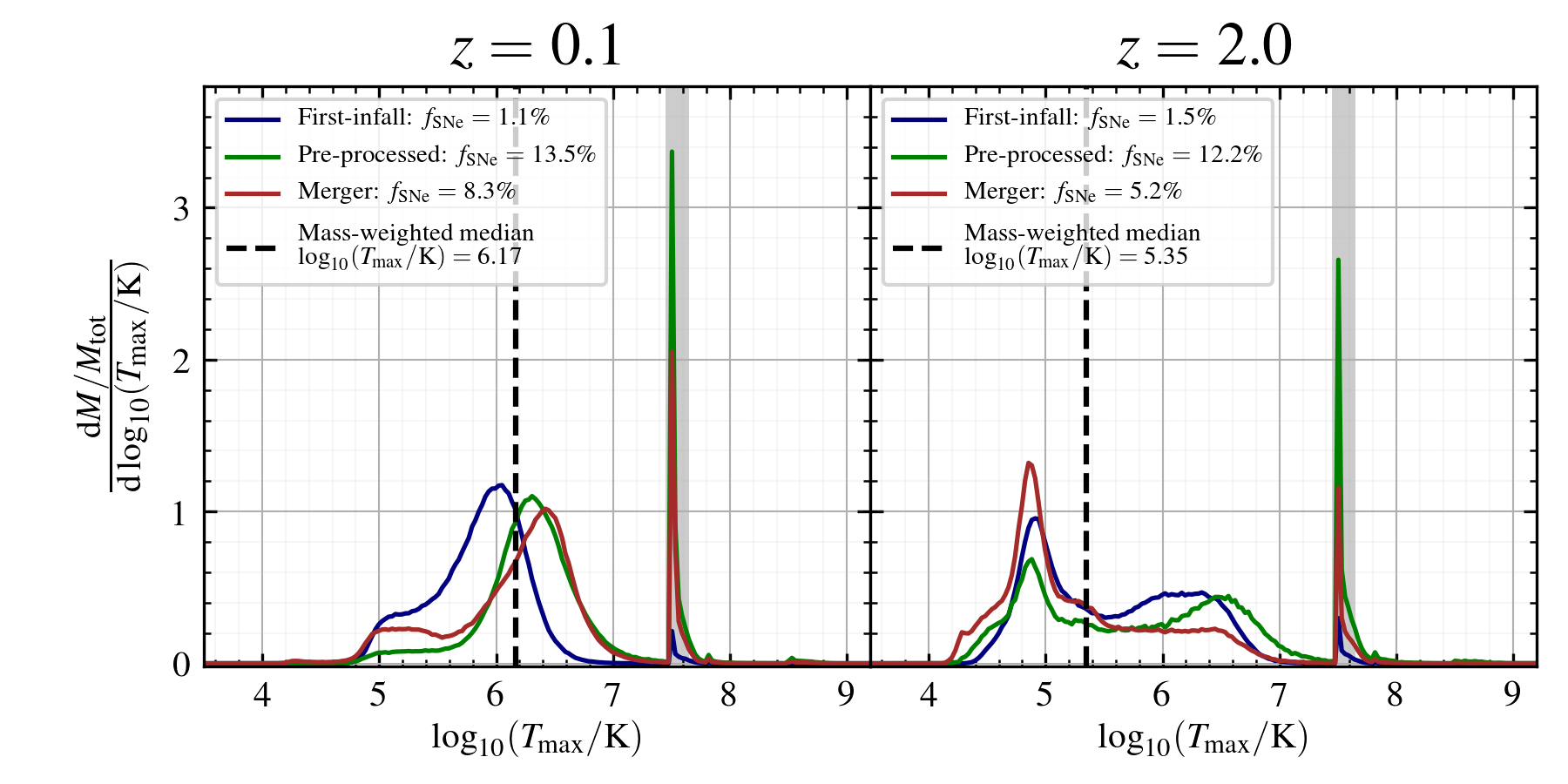}
    \caption{The mass-weighted PDF of pre-accretion gas $T_{\rm max}$ values in L50-REF, split by history-based accretion mode. The PDF is shown for $z\approx0$ in the left panel, and $z\approx2$ in the right panel, and includes all gas particles accreting to haloes in the mass range $10^{12}M_{\odot}<M_{\rm halo}<10^{12.5}M_{\odot}$. These distributions reflect the maximum temperatures that gas particles had reached prior to accretion. We also included the mass-weighted median $T_{\rm max}$ values as dashed lines. We quantify the amount of gas heated by stellar feedback by identifying gas particles with $T_{\rm max}$ values in a band of $0.1$~dex close to the $\Delta T_{\rm SNe}$ value of $10^{7.5}{\rm K}$, quoted in the legend.}
    \label{fig:apdx:tmax:channels}
\end{figure*}
Figure \ref{fig:apdx:tmax:channels} illustrates the mass-weighted PDF of the maximum temperature gas particles reached prior to accretion at $z\approx0$ (left panel) and $z\approx2$ (right panel) in L50-REF. We use haloes in the mass band between $10^{12}M_{\odot}$ and $10^{12.5}M_{\odot}$, and split the accreting particles based on their inflow channel. We quantify the amount of gas heated by stellar feedback by identifying gas particles with $T_{\rm max}$ values in a band of $0.1$~dex close to the $\Delta T_{\rm SNe}$ value of $10^{7.5}{\rm K}$, and quote this proportion as a fraction of mass in the legend. 

Encouragingly, we find that the first-infall mode has a very small proportion of contamination by stellar feedback at $\approx1-1.5\%$ for both redshift selections. In comparison, $12-13\%$ of pre-processed accreting gas mass has been directly affected by stellar feedback. We believe the slight contamination in the first-infall mode could be a result of the simulation cadence - if a particle was accreted onto a halo and swiftly ejected within the gap between simulation outputs, our algorithm would not identify the particle as ``processed''. 

The significant proportion of pre-processed and merger gas with $T_{\rm max}$ values near the $\Delta T_{\rm SNe}$ value highlights the value in separating inflow modes based on current temperature (as in \citealt{Correa2018}) as opposed to their $T_{\rm max}$ value. Using a  $T_{\rm max}$ cutoff would not allow particles to cool and does not reflect the extent of virial shock-heating which we are attempting to quantify.

\section{Convergence of covering fractions}\label{sec:appendix:fcov}

\begin{figure*}
    \centering
    \includegraphics[width=0.9\textwidth]{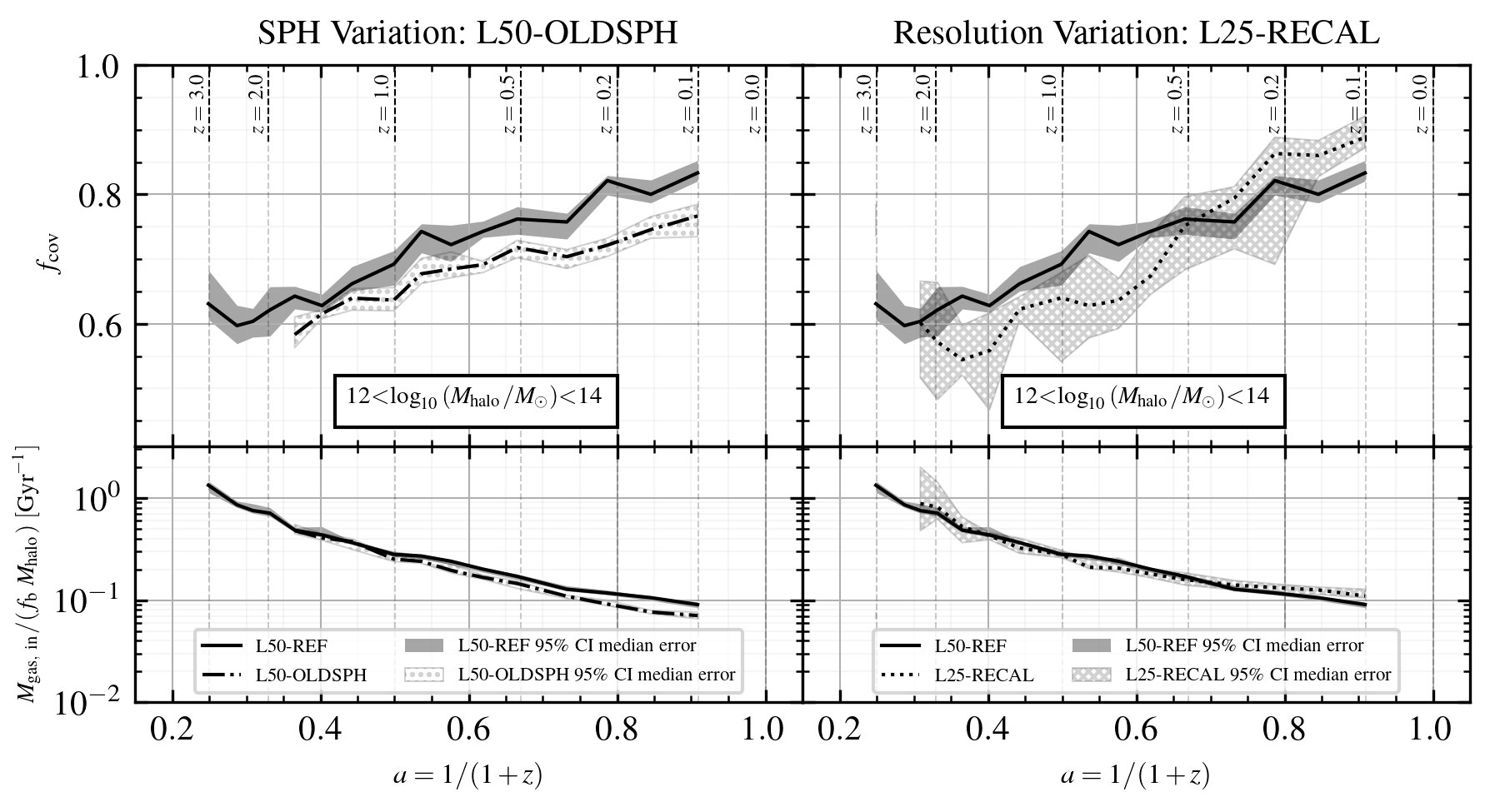}
    \caption{Top panels: The median covering fraction, $f_{\rm cov}$, of gas accreting to selected \eagle\ haloes as a function of the scale factor, $a=1/(1+z)$, comparing L50-REF with L50-OLDSPH in the left panel and L50-REF with L25-RECAL in the right panel. We note this $f_{\rm cov}$ in this plot is not broken down into different modes of accretion, but is calculated for all accreting particles regardless of their history or temperature. Haloes are included at each snap if they are (i) within the mass range $10^{12}M_{\odot}\lesssim M_{\rm halo}\lesssim10^{14}M_{\odot}$, and (ii) have accreted $\geq10^{3}$ gas particles since the last snapshot. Bottom panels: the median halo-scale gas accretion efficiency (using the same halo selections as above) as a function of scale factor, comparing L50-REF with L50-OLDSPH in the left panel and L50-REF with L25-RECAL in the right panel. In each panel, we also include the bootstrap-generated $95\%$ confidence interval on the median from 100 resamples of the median, with half of the respective populations. }
    \label{fig:apdx:res:fcov}
\end{figure*}

In Figure \ref{fig:apdx:res:fcov}, we compare total covering fractions and accretion efficiencies of haloes (in the mass range $10^{12}M_{\odot}-10^{14}M_{\odot}$) in the L50-OLDSPH and L25-RECAL runs to the reference physics run (L50-REF). This allows us to measure the influence of SPH implementation and mass resolution respectively on values of $f_{\rm cov}$  (defined in Equation \ref{eq:fcov}) and total accretion rates. 

The L50-OLDSPH \eagle\ run (see \citealt{Schaller2015}) uses an older density-SPH formulation, which is known for its weakness at modelling discontinuities and mixing (e.g. \citealt{Agertz2007}). Given the established mixing problems and expectation for more ``clumpy'' structure formation in this run, we compare the covering fraction of inflow between this run and the L50-REF run in the top left panel of Figure \ref{fig:apdx:res:fcov}. We find that across cosmic time, gas accretion onto haloes is marginally (but significantly) more collimated than accreting gas in L50-REF by $\approx5-10\%$. Over the same time interval, we find that haloes in the relevant mass range show similar total accretion efficiencies (bottom left panel). Interestingly, despite the differences we find in the spatial distribution of accreting gas, \citet{Schaller2015} show that most galaxy properties are not heavily influenced by the choice of SPH solver. This indicates that it is total gas accretion rate (and not necessarily its spatial characteristics) that plays the dominant role in shaping halo and galaxy properties in cosmological-scale simulations.

\citet{Schaye2015} introduced the concept of ``strong'' and ``weak'' convergence tests. Strong convergence refers to the case where a simulation is re-run at higher resolution, with both better mass and spatial resolution, adopting exactly the same sub-grid physics models and parameters. Weak convergence refers to the case when a simulation is re-run at higher resolution but the sub-grid parameters are recalibrated to recover, as best as possible, the level of agreement with the adopted calibration diagnostic (in the case of \eagle, the $z=0.1$ galaxy stellar mass function and stellar size-mass relation of galaxies). With this purpose, two higher-resolution versions of \eagle\ were introduced by \citet{Schaye2015} - both in a box of ($25$~cMpc)$^{3}$ with $2\times 752^3$ particles. These simulations have better mass and spatial resolution than the intermediate-resolution of the L50-REF simulation by factors of $8$ and $2$, respectively. 

We check the weak convergence of our $f_{\rm cov}$ parameter by using the L25N752-RECAL run (top right panel, Figure \ref{fig:apdx:res:fcov}), with $4$ recalibrated parameters that were tuned to reproduce the $z=0$ observables above. In this mass range we see that within the bootstrap-generated uncertainty on the median $f_{\rm cov}$ value over redshift, the covering fraction of accreting gas in L25-RECAL and L50-REF are largely consistent. At high redshift, $z\gtrsim1$, covering fractions in L25-RECAL may be marginally lower than we find in L50-REF, however it is difficult to ascertain the significance of this disparity with the limited number of sufficiently sized haloes in L25-RECAL. We thus conclude that covering fractions appear to be consistent between L25-RECAL and L50-REF, indicating that there is weak convergence in the $f_{\rm cov}$ parameter.

\bsp	
\label{lastpage}
\end{document}

%% file: Tables/Simulation-table.tex
\begin{tabular}{|c||c|c|c|c|l|c|c|c|c|}
\hline
Run Name   & $L_{\rm box}$/cMpc & $N_{\rm part}$ & $m_{\rm DM}/{\rm M}_{\odot}$ & $m_{\rm gas}/{\rm M}_{\odot}$ & {Anarchy SPH} & $\epsilon$/pkpc & $\Delta T_{\rm SNe
}/{\rm K}$ & $\Delta T_{\rm AGN}/{\rm K}$ & $N_{{\rm halo},\ z=0}\ (>10^9M_{\odot})$ \\ \hline \hline
L50-REF    & 50                       & $752^3$        & $9.7\times10^6$            & $1.8\times10^6$             & {\checkmark}                   & $0.70$                    & $10^{7.5}$                      & $10^{8.5}$                       & $74,743$                                 \\ \hline
L50-NOAGN  & 50                       & $752^3$        & $9.7\times10^6$            & $1.8\times10^6$             & {\checkmark}                   & $0.70$                    & $10^{7.5}$                      & N/A                              & $74,351$                                 \\ \hline
L25-NOFB   & 25                       & $376^3$        & $9.7\times10^6$            & $1.8\times10^6$             & {\checkmark}                   & $0.70$                    & N/A                             & N/A                              & $9,362$                                  \\ \hline
L50-OLDSPH & 50                       & $752^3$        & $9.7\times10^6$            & $1.8\times10^6$             & {$\times$}                     & $0.70$                    & $10^{7.5}$                      & $10^{8.5}$                       & $77,741$                                 \\ \hline
L25-RECAL  & 25                       & $752^3$        & $1.2\times10^6$            & $2.3\times10^5$             & {\checkmark}                  & $0.35$                    & $10^{7.5}$                      & $10^{9}$                         & $10,068$                                 \\ \hline
\end{tabular}

%% file: Tables/Classification-table.tex
\begin{tabular}{|c||c|l|c|}
\hline
Inflow channel          & Classification type  & \multicolumn{1}{c|}{Description}                                                               & Colour                                          \\ \hline \hline
Total accretion         & N/A                  & Particles identified as accreted: not part of FOF at snap $n-1$ but appear in FOF at snap $n$. & \cellcolor[HTML]{C0C0C0}                        \\ \hline
First-infall accretion  & Particle history     & Particles identified as accreted from the field and never previously processed in a halo.      & \cellcolor[HTML]{00009B}{\color[HTML]{303498} } \\ \hline
Pre-processed accretion & Particle history     & Particles identified as accreted from the field and previously processed in a halo.            & \cellcolor[HTML]{32CB00}                        \\ \hline
Merger accretion        & Particle history     & Particles identified as accreted that were in a separate halo at snap $n-1$.                   & \cellcolor[HTML]{B70A0A}{\color[HTML]{CB0000} } \\ \hline
Hot accretion           & Particle temperature & Particles identified as accreted from the field with snap $n$ temperature above $10^{5.5}$ K.  & \cellcolor[HTML]{F37EE3}{\color[HTML]{DAE8FC} } \\ \hline
Cold accretion          & Particle temperature & Particles identified as accreted from the field with snap $n$ temperature below $10^{5.5}$ K.  & \cellcolor[HTML]{ABCBF8}                        \\ \hline
\end{tabular}